\documentclass[journal=jacsat,manuscript=article]{achemso}
\setkeys{acs}{maxauthors=99}
\usepackage[utf8]{inputenc}
\usepackage{physics}
\usepackage{braket}
\usepackage{bm}%
\usepackage[hidelinks]{hyperref}%
\SectionNumbersOn 
\usepackage{multirow}
\usepackage{xcolor}
\usepackage{graphicx}
\usepackage{qcircuit}
\usepackage{tikz}
\usetikzlibrary{positioning}
\usepackage{enumitem}
\setlist{nosep}
\usepackage{courier}
\usepackage[version=4]{mhchem}

\usepackage{listings}
\definecolor{codegreen}{rgb}{0,0.6,0}
\definecolor{codegray}{rgb}{0.5,0.5,0.5}
\definecolor{codepurple}{rgb}{0.58,0,0.82}
\definecolor{backcolour}{rgb}{0.95,0.95,0.92}

\lstdefinestyle{mystyle}{
    backgroundcolor=\color{backcolour},   
    commentstyle=\color{codegreen},
    keywordstyle=\color{magenta},
    numberstyle=\tiny\color{codegray},
    stringstyle=\color{codepurple},
    basicstyle=\linespread{0.9}\ttfamily\footnotesize,
    breakatwhitespace=false,         
    breaklines=true,                 
    captionpos=b,                    
    keepspaces=true,                 
    numbers=left,                    
    numbersep=5pt,                  
    showspaces=false,                
    showstringspaces=false,
    showtabs=false,                  
    tabsize=2,
    basewidth=0.55em,
}

\newcommand{\bp}[1]{\vspace{\baselineskip}\noindent\textbf{#1}}
\definecolor{bkgd}{RGB}{240,242,246}
\definecolor{orange-red}{rgb}{1.0, 0.27, 0.0}
\newcommand{\fancylink}[2]{\colorbox{bkgd}{\color{orange-red}\href{#1}{\sf {#2}}}}

\newcommand{\tc}{\textsc{TensorCircuit}}
\newcommand{\tcc}{\textsc{TenCirChem}}
\newcommand{\jax}{\textsc{JAX}}
\newcommand{\numpy}{\textsc{NumPy}}
\newcommand{\cupy}{\textsc{CuPy}}
\newcommand{\pyscf}{\textsc{PySCF}}
\newcommand{\scipy}{\textsc{SciPy}}
\newcommand{\reno}{\textsc{Renormalizer}}
\newcommand{\openfermion}{\textsc{OpenFermion}}

\newcommand{\qiskit}{\textsc{QisKit}}

\newcommand{\pennylane}{\textsc{PennyLane}}
\newcommand{\tequilla}{\textsc{Tequilla}}
\newcommand{\mindquantum}{\textsc{MindQuantum}}
\newcommand{\qschem}{\textsc{Q$^2$ Chemistry}}
\newcommand{\qforte}{\textsc{QForte}}
\newcommand{\nex}{N_{\rm{ex}}}
\newcommand{\nshots}{N_{\rm{shots}}}

\usepackage{cancel}
\newcommand{\fix}[1]{\textcolor{red}{#1}}
\renewcommand{\cancel}[1]{}
\renewcommand{\fix}[1]{#1}

\title{TenCirChem: An Efficient Quantum Computational Chemistry Package for the NISQ Era}

\author{Weitang Li}
\email{liw31@gmail.com}
\affiliation{Tencent Quantum Lab, Shenzhen, 518057, China}

\author{Jonathan Allcock}
\affiliation{Tencent Quantum Lab, Hongkong, 999077, China}

\author{Lixue Cheng}
\affiliation{Tencent Quantum Lab, Shenzhen, 518057, China}

\author{Shi-Xin Zhang}
\affiliation{Tencent Quantum Lab, Shenzhen, 518057, China}

\author{Yu-Qin Chen}
\affiliation{Tencent Quantum Lab, Shenzhen, 518057, China}

\author{Jonathan P. Mailoa}
\affiliation{Tencent Quantum Lab, Shenzhen, 518057, China}

\author{Zhigang Shuai}
\affiliation{Department of Chemistry, Tsinghua University, Beijing, 100084, China}
\alsoaffiliation{School of Science and Engineering, The Chinese University of Hong Kong, Shenzhen, 518172, China}

\author{Shengyu Zhang}
\email{shengyzhang@tencent.com}
\affiliation{Tencent Quantum Lab, Hongkong, 999077, China}

\begin{document}

\begin{abstract}
    \tcc{} is an open-source Python library for simulating variational quantum algorithms for quantum computational chemistry.
    \fix{
     \tcc{} shows high-performance on the simulation of unitary coupled-cluster circuits, using compact representations of quantum states and excitation operators. Additionally, \tcc{} supports noisy circuit simulation and provides algorithms for variational quantum dynamics. \tcc{}'s capabilities are demonstrated through various examples, such as the calculation of the potential energy curve of \ce{H2O} with a 6-31G(d) basis set using a 34-qubit quantum circuit, the examination of the impact of quantum gate errors on the variational energy of the \ce{H2} molecule, and the exploration of the Marcus inverted region for charge transfer rate based on variational quantum dynamics.
    Furthermore, \tcc{} is capable of running real quantum hardware experiments, making it a versatile tool for both simulation and experimentation in the field of quantum computational chemistry.
    }
\end{abstract}

\maketitle

\section{Introduction}
Quantum computation leverages quantum effects to store and process data, which could lead to a revolution in computational chemistry~\cite{mcclean2016theory, cao2019quantum, bauer2020quantum, mcardle2020quantum, liu2022quantum}.
\fix{\cancel{As the era of} With the advent of noisy intermediate scale quantum (NISQ) computing~\cite{Preskill18} \cancel{gathers steam}}, devices with tens \fix{\cancel{to hundreds}} of error-prone qubits are increasingly becoming available \fix{for use} to researchers and the public~\cite{arute2019quantum,gong2021quantum, xu2023digital, google2023suppressing}.  Since such devices are not capable of running large-depth structured quantum algorithms, the question of how to best utilize them to solve chemistry problems is \fix{\cancel{one}} of deep scientific and commercial significance.
In recent years, following pioneering proposals such as the variational quantum eigensolver (VQE)~\cite{peruzzo2014variational}, research has focused on variational algorithms based on parameterized quantum circuits (PQC)~\cite{o2016scalable, cerezo2021variational, tilly2022variational}. 
In the long term, the quantum phase estimation (QPE) algorithm offers a promising route to evaluating the energy of molecular systems, as it may potentially provide an exponential advantage over classical full configuration interaction (FCI) treatment~\cite{aspuru2005simulated, lee2023evaluating}.
However, compared with VQE, the circuit depths required to run QPE are typically far too deep to run on near-term quantum devices, making VQE the approach of choice in the NISQ era.

VQE uses a parameterized quantum circuit to represent a system's wavefunction and works by tuning the circuit parameters to minimize the energy variationally.
In this setting, the quantum circuit plays the same role as the ansatz in classical quantum chemistry algorithms and, similarly, a variety of different \fix{\cancel{ansatze}ans\"{a}tze} have been proposed.
The first such proposal was the \fix{disentangled} unitary coupled cluster (UCC) ansatz~\cite{peruzzo2014variational}, which is a variant of the classical coupled cluster method.
\fix{While disentangled UCC differs from the traditional UCC ansatz developed for classical computers~\cite{bartlett1989alternative, kutzelnigg1991error}, 
for brevity, in the following, we will omit the word ``disentangled''. }
UCC is known to be difficult to implement in classical computers, yet the ansatz can be transformed into a PQC straightforwardly~\cite{anand2022quantum} and implemented on a quantum computer.
While excitations are commonly restricted to single and double excitations -- a case referred to as UCCSD -- a number of other variants of UCC have also been proposed~\cite{lee2018generalized, grimsley2019adaptive, elfving2021simulating}.
Another popular family of \fix{\cancel{ansatze}ans\"{a}tze} is the hardware-efficient ansatz (HEA)~\cite{kandala2017hardware}, which is designed to be first and foremost
easily implementable on a target quantum hardware platform, and includes fewer quantum chemistry heuristics compared to UCC.
In addition to ground state properties\fix{,} such as molecular energy, chemists are also interested in quantum dynamics~\cite{ollitrault2021molecular}.
Under the PQC framework, the time evolution of the circuit parameters can be determined using time-dependent variational principles~\cite{li2017efficient, yuan2019theory}. 

While the quality and availability of quantum computing hardware continue to improve, the majority of real experiments are currently still constrained to small numbers of qubits and very \fix{\cancel{short}shallow} circuits \fix{\cancel{depths}}~\cite{kandala2017hardware, colless2018computation, kandala2019error, rice2021quantum, gao2021applications, kirsopp2022quantum}.  
\fix{
Furthermore, optimizing circuit parameters on real devices is challenging: gradient-based methods are inefficient, as they rely on computing finite differences or parameter-shifts~\cite{mitarai2018quantum, schuld2019evaluating}, and both gradient and gradient-free methods are susceptible to noise.
}
These factors mean that, for the foreseeable future, the classical simulation of variational quantum algorithms will play an important role in studying and analyzing their performance and viability.  Although a variety of simulation software is currently available, there remain opportunities to significantly improve \fix{efficiency, functionality, and flexibility}.

Here we introduce \tcc{}, an \fix{efficient} open-source Python library for simulating variational quantum chemistry algorithms, designed to be easy to use as a black box while allowing for a high degree of flexibility and customizability.
\tcc{} is powered by \tc{}~\cite{zhang2023tensorcircuit}, a recently released quantum circuit simulation package featuring an advanced tensor network contraction simulation engine.
A recently developed algorithm to efficiently simulate the UCC ansatz on classical computers~\cite{chen2021quantum, kottmann2021feasible, rubin2021fermionic} is also implemented in \tcc{} with notable performance optimizations.
We demonstrate how to use \tcc{} by providing concrete code snippets, as well as links to online tutorials for features\fix{,} such as efficient UCC calculation, noisy circuit simulation, and quantum dynamics simulation.
In the examples, we include an accurate UCCSD potential energy curve of the water molecule corresponding to 34 qubits, the influence of quantum gate error rate and measurement shots on VQE energy, the spin relaxation dynamics of the spin-boson model, and much more.

The paper is structured as follows.
In the rest of the section, we briefly review variational quantum algorithms for quantum chemistry and existing simulation packages.
In Sec.~\ref{sec:overview}, we present the overall architecture and workflow of \tcc{} with an emphasis on \fix{\cancel{their}related} theoretical background.
Secs.~\ref{sec:ucc}, \ref{sec:hea}, and \ref{sec:dynamics} illustrate the features contained in the \tcc{} modules through code snippets and detailed simulation examples.

\subsection{Variational quantum algorithms for quantum chemistry}
\label{sec:intro-vqa}
Variational quantum algorithms for quantum chemistry have been an active area of research since the original VQE proposal~\cite{peruzzo2014variational}.
For both ground state properties and quantum dynamics, 
while details vary, proposed protocols typically adhere to the following workflow:

\bp{Input.} The input to the problem is the system Hamiltonian in second-quantized form.
For electronic structure problems, this can be expressed as
\begin{equation}
\label{eq:ham-abinit}
    H = \sum_{pq}h_{pq}a^\dagger_p a_q + \fix{\frac{1}{2}}\sum_{pqrs}h_{pqrs}a^\dagger_p a^\dagger_q a_r a_s + E_{\rm{nuc}}, \
\end{equation}
where $h_{pq}$ and $h_{pqrs} = [ps|qr]$ are one-electron and two-electron integrals, and $a^\dagger_p, a_p$ are fermionic creation and annihilation operators, respectively, acting on the $p$-th spin-orbital.
$E_{\rm{nuc}}$ is the nuclear repulsion energy.
For quantum dynamics simulations, the Hamiltonian varies from system to system,
and may contain only electronic terms~\cite{lee2021simulation, lee2022simulating} 
or both electronic and vibrational terms~\cite{ollitrault2020nonadiabatic, lee2022variational}. 
The latter is referred to as vibronic-coupling Hamiltonians.

\bp{Conversion to qubits.} The Hamiltonian, either electronic or vibronic-coupling,
is converted into an $N$-qubit Hamiltonian of the form
\begin{equation}
\label{eq:ham-pauli}
    H = \sum_j^M \alpha_j P_j, \
\end{equation}
where the $\alpha_j$ are real coefficients, $M$ is the total number of terms, and the $P_j$ are Pauli strings of the form $P_j = \sigma_{i_1}\otimes \sigma_{i_2}\otimes\ldots \sigma_{i_N}$ where the \fix{$\sigma_{i_k}$} are single qubit Pauli operators or the identity. 
For electronic Hamiltonians, the Jordan-Wigner transformation~\cite{JW28}, parity transformation~\cite{bravyi2002fermionic, seeley2012bravyi} and Bravyi-Kitaev transformation~\cite{seeley2012bravyi} are popular choices to perform this conversion.
The Jordan-Wigner and Bravyi-Kitaev transformations require one qubit per spin-orbital, 
while the parity transformation is able to save 2 qubits, relative to these other methods, due to electron number conservation in each spin sector.
For vibronic-coupling Hamiltonians, the nuclear states also need to be encoded into qubits via unary or binary encodings~\cite{sawaya2020resource}. 

\bp{Choice of ansatz.} A parameterized ansatz state $\ket{\psi(\theta)} = U(\theta)\ket{\psi_0}$ is selected, where $U(\theta)$ is the unitary corresponding to a quantum circuit characterized by a vector $\theta$ of tunable parameters, and $\ket{\psi_0}$ is a chosen reference state or initial state.
The core role of quantum computers \fix{or quantum processing units} (QPUs) in variational algorithms is to use quantum circuits to execute $U(\theta)$ and prepare $\ket{\psi(\theta)}$.
Thus, the ability of the variational ansatz in expressing the system wavefunction is a key factor
impacting the accuracy of variational quantum algorithms for quantum chemistry.
For electronic Hamiltonians, the proposed \fix{\cancel{ansatze}ans\"{a}tze} generally fall into unitary coupled cluster (UCC) based approaches~\cite{peruzzo2014variational, grimsley2019adaptive, nam2020ground, li2022toward, guo2022experimental}, 
and hardware-efficient \fix{\cancel{ansatze}ans\"{a}tze}~\cite{kandala2017hardware,colless2018computation, kandala2019error, rice2021quantum, gao2021applications, kirsopp2022quantum}.
For quantum dynamics simulation,
the appropriate family of ansatz varies from system to system.
The variational Hamiltonian ansatz~\cite{wecker2015progress, miessen2021quantum} and adaptive ansatz~\cite{yao2021adaptive} \fix{automate the task of constructing problem-specific circuit ans\"{a}tze}.

\bp{Updating parameters.} In the case of VQE, 
\fix{
because the true ground state energy $E_0$ of the system gives a lower bound on the energy expectation value $\langle H \rangle_\theta := \bra{\psi(\theta)}H \ket{\psi(\theta)}$, $\langle H \rangle_\theta$ is minimized with respect to the parameters $\theta$, 
}
which \fix{in turn} gives an upper bound on $E_0$ i.e.,
\begin{align*}
    E_0 &\le \min_{\theta}\langle H \rangle_\theta.
\end{align*}
The quality of this bound depends \fix{\cancel{not only}} on the chosen ansatz. In general, the function is non-convex and one can only expect to find a local minimum, which depends on the parameter initializations and the optimization protocol used.
For dynamics problems, the parameters are propagated according to time-dependent variational principles,
of which McLachlan's variational principle 
is among the most popular choices for quantum computing~\cite{broeckhove1988equivalence, yuan2019theory}.
McLachlan's variational principle minimizes the distance between the ideal time derivative $-iH\ket{\psi(\theta)}$, i.e.
and actual time derivative $\pdv{}{t}\ket{\psi(\theta)}$
\begin{equation}
    \min_{\dot \theta} \norm{\left(\pdv{}{t} + iH\right )\ket{\psi(\theta)}}. \
\end{equation}
At each time step, $\dot \theta = \pdv{\theta}{t}$ is solved and then $\theta$ is evolved numerically according to $\dot \theta$.
The equation of motion for $\theta_k$, the $k$th element in vector $\theta$, is
\begin{equation}
\label{eq:eom-theta}
\sum_k M_{jk} \dot \theta_k =  V_j  , \
\end{equation}
where
\begin{equation}
\label{eq:dynamic-mv}
    M_{jk} = \Re{ \pdv{\bra{\psi}}{\theta_j} \pdv{\ket{\psi}}{\theta_k}  } , \
    V_j = \Im{  \pdv{\bra{\psi}}{\theta_j}  H \ket{\psi} }
\end{equation}
Both $M_{jk}$ and $V_j$
can be evaluated on a quantum computer with one additional ancilla qubit via the Hadamard test~\cite{li2017efficient}.
$\dot \theta_k$ can then be deduced using a standard linear equation solver on a classical computer.
After that, $\theta_k$ is propagated in time with time step $\tau$ either by the simple forward Euler method
$\theta_k(t + \tau) = \theta_k(t) + \tau \dot \theta_k(t)$
or more sophisticated initial value problem solvers\fix{,} such as Runge-Kutta solvers.

\subsection{Existing packages: Challenges and opportunities} 
A number of useful software packages are currently available, which facilitate the research, design, and validation of quantum computational chemistry algorithms.
A variety of general quantum computing packages, such as \qiskit{}~\cite{qiskit}, \pennylane{}~\cite{bergholm2018pennylane}, and \mindquantum{}~\cite{mq_2021}, have modules or sub-packages dedicated to chemistry applications, and there are also standalone quantum simulation packages that are mostly or completely 
designed for quantum computational chemistry, such as \tequilla{}~\cite{kottmann2021tequila}, \qschem{}~\cite{fan2022q}, and \qforte{}~\cite{stair2022qforte}. 

The efficiency and scalability of the packages depend crucially on the underlying circuit simulation algorithm.
The most commonly encountered quantum circuit simulator is statevector simulator, which is implemented in all simulation packages mentioned above.
By using statevector simulator, all (exponentially many with respect to the number of qubits) amplitudes of the wavefunction are computed and stored. For such simulators, the largest simulated UCC circuit to date contains fewer than 30 qubits~\cite{cao2022progress}.
In contrast, \fix{\cancel{M}m}atrix product state (MPS) simulators are able to tackle larger circuits, with \fix{around 100 qubits~\cite{schollwock2011density, shang2023towards}} possible.
Among the packages mentioned above,  \qiskit{}, \pennylane{}, and \qschem{} have implemented the MPS simulator.
However, MPS simulation involves approximations which could be problematic for the validation of quantum algorithms.
Another challenge for software development is noisy circuit simulation and interfacing with hardware.
\fix{
These features can help researchers better understand the behavior of quantum algorithms in real-world conditions, facilitating the development of noise-resilient quantum algorithms and better error mitigation techniques~\cite{ollitrault2020hardware, miessen2021quantum, li2022toward}, which are essential for building practical quantum computers.}
\qiskit{} is currently the only package capable of noisy circuit simulation \fix{and interfacing with quantum hardware. \cancel{and}Furthermore,} to the best of the authors' knowledge, quantum dynamics algorithms of molecular systems are not implemented in any packages currently available.

\tc{} is a recent open-source quantum circuit simulator based on tensor network contraction which -- in certain cases -- enables the simulation of up to 600 qubits without any approximation~\cite{zhang2023tensorcircuit}.
Differences between tensor network contraction simulators and MPS simulators are described in Sec.~\ref{sec:overview-engines}.
\tc{} is built upon industry-leading machine learning frameworks\fix{,} such as \jax{}~\cite{jax2018github}
and employs automatic differentiation (AD),
just-in-time (JIT) compilation, and vectorized parallelism to accelerate simulation.
\tc{} also possesses a fully-fledged noise model which enables efficient noisy circuit simulation. 
With its \tc{} backend, 
\tcc{} provides a competitive option for developing and analyzing quantum computational chemistry algorithms.
\fix{The role of \tc{} in \tcc{} is described in more detail in Sec.~\ref{sec:archi-workflow}.}

\section{TenCirChem Overview and Theoretical Backgrounds}
\label{sec:overview}

In this section, we provide a high-level picture of \tcc{} by introducing its architecture and typical workflow. In addition, we present the theoretical background relevant to the code examples found in Secs.~\ref{sec:ucc}, \ref{sec:hea} and \ref{sec:dynamics}.

\subsection{Architecture and workflow}
\label{sec:archi-workflow}
\begin{figure}[htp]
    \centering
    \includegraphics[width=1.0\textwidth]{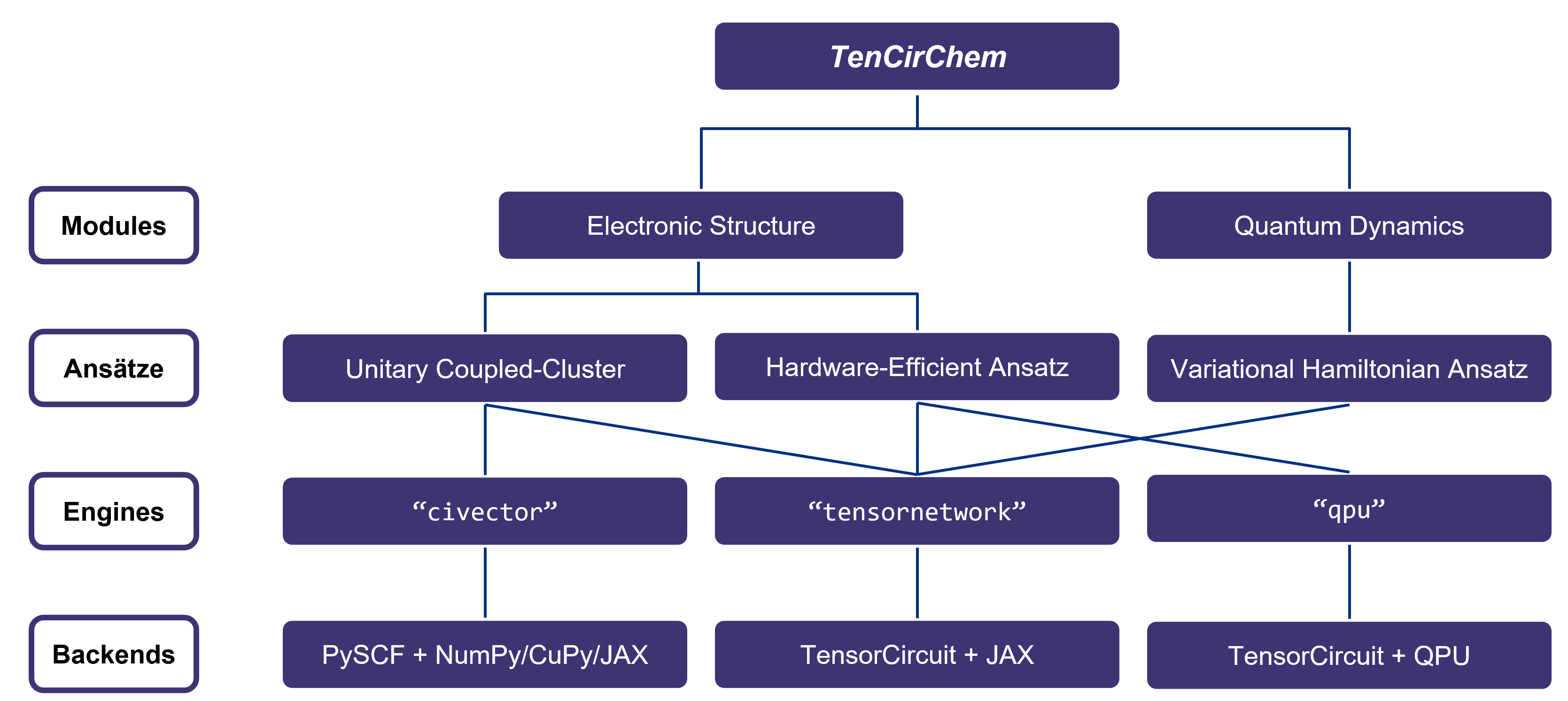}
\caption{The architecture of \tcc{}. The electronic structure and quantum dynamics modules are described in depth in Secs.~\ref{sec:ucc}, \ref{sec:hea} and \ref{sec:dynamics}. See Sec.~\ref{sec:supported-ansatz} for more information on the supported ans\"{a}tze. For more details on engines and backends see Sec.~\ref{sec:overview-engines}.}
    \label{fig:architecture}
\end{figure}

\bp{Architecture.} \tcc{}
consists of two primary modules (Fig.~\ref{fig:architecture}), corresponding to (i) electronic structure and (ii) quantum dynamics.
Both modules contain a set of pre-built \fix{\cancel{ansatze}ans\"{a}tze}, and user-specified custom \fix{\cancel{ansatze}ans\"{a}tze} are also supported.
Due to the popularity and unique mathematical structure of the UCC ansatz, an efficient simulation engine \lstinline{"civector"}
is implemented specifically for its efficient simulation. The details of the engine are included in the Supporting Information.
In short, the engine exploits particle number conservation to store the system wavefunction in \textit{configuration interaction vector} (\lstinline{"civector"}) form and uses \textit{UCC factor expansion} to accelerate simulation.
This engine relies heavily on \fix{the FCI module of} \pyscf{}~\cite{sun2018pyscf} \fix{for the construction and manipulation of the configuration interaction vector}.
\fix{
In addition, the gradient is evaluated by auto-differentiation with reversible computing, which is more memory-efficient than the traditional reverse-mode auto-differentiation implemented in standard machine-learning packages~\cite{luo2020yao}.
}
The \lstinline{"tensornetwork"} engine uses the \tc{} tensor network contraction kernels to perform circuit simulation. \fix{A variant of the \lstinline{"tensornetwork"} engine is } the \lstinline{"tensornetwork-noise"} engine\fix{, which} further includes circuit gate noise and measurement uncertainty in the simulation using the relevant utilities implemented in \tc{}.
\fix{The \lstinline{"qpu"} engine delegates circuit execution to real QPUs through \tc{}.}

\bp{Workflow.} As mentioned in the introduction, the starting point for variational quantum chemistry algorithms is normally taken to be the Hamiltonian in the second-quantized form.  For electronic structure problems, this means specifying the one- and two-electron integrals $h_{pq}$ and $h_{pqrs}$ in Eq.~\ref{eq:ham-abinit}. 
In \tcc{} this can be done either directly, by inputting these numbers explicitly, or indirectly, by specifying the 3D coordinates, atom types, and basis sets of each of the atoms in a molecule and having \pyscf{} compute the integrals automatically.  
The \fix{\cancel{general}} Hamiltonian \fix{\cancel{met}} in quantum dynamics is defined using the \reno{} \cite{renormalizer} package by specifying each term in the Hamiltonian.

Next, the Jordan-Wigner, parity \fix{or Bravyi-Kitaev conversion} from electronic Hamiltonian (Eq.~\ref{eq:ham-abinit}) to qubit Hamiltonian (Eq.~\ref{eq:ham-pauli}) is performed by \openfermion{}~\cite{mcclean2020openfermion}. 
The encoding of nuclear states to qubits and the construction of the \fix{\cancel{ansatze}ans\"{a}tze} are performed within the \tcc{} package.
The workflow and the underlying packages are summarized in Fig.~\ref{fig:workflow}\fix{, 
in accordance with Sec.~\ref{sec:intro-vqa}}, and will be elaborated on in the following sections.
Standard infrastructure packages, such as \numpy{} and \scipy{}, are omitted for brevity.
\fix{The specific tasks of each underlying package are summarized below:
\begin{itemize}
    \item \pyscf{}: specifying molecules; calculating integrals; configuration interaction vector construction and manipulation.
    \item \openfermion{}: specifying operators in fermion operator or qubit operator format; conversion from fermion operator to qubit operator.
    \item \reno{}: specifying general quantum dynamics Hamiltonian; conversion from symbolic operators to numeric operators.
    \item \tc{}: constructing custom PQC as ansatz; quantum circuit simulation via tensor network contraction; interfacing with QPU.
\end{itemize}
}

\begin{figure}
\centering
\tikzstyle{base} = [rectangle, rounded corners, thick,
scale=0.85,
minimum width=4cm, 
text width=3.5cm,
minimum height=2cm,
text centered]
\tikzstyle{comment} = [rectangle, rounded corners, thick,
scale=0.85,
minimum height=2cm]
\tikzstyle{task} = [base, draw=black]
\tikzstyle{arrow} = [very thick,->,>=stealth]
\begin{tikzpicture}[node distance=2.5cm]

\node (n1) [task]  {Specify Hamiltonian};
\node (n2) [task, xshift=5cm]  {Convert to qubits};
\node (n3) [task, xshift=10cm]  {Construct ansatz};
\node (n4) [task, xshift=15cm]  {Simulate circuit and update parameter};

\node (m1) [base, below of=n1] {\textcolor{black}{\pyscf{}$^{a}$} \\  \textcolor{black}{\openfermion{}$^{a}$} \\ \textcolor{black}{\reno{}$^{b}$} };
\node (m2) [base, below of=n2] {\textcolor{black}{\openfermion{}$^{a}$} \\ \textcolor{orange}{\tcc{}$^{b}$}};
\node (m3) [base, below of=n3] {\textcolor{orange}{\tcc{}$^{a, b}$}};
\node (m4) [base, below of=n4] {\textcolor{orange}{\tcc{}$^{a, b}$} \\ \textcolor{black}{\tc{}$^{a, b}$} \\ \textcolor{black}{\pyscf{}$^{a}$}};

\node (m1) [comment, below=1.5cm of m1.west, anchor=west]  {$^{a}$For electronic structure. $^{b}$For quantum dynamics.};

\draw [arrow] (n1) -- (n2);
\draw [arrow] (n2) -- (n3);
\draw [arrow] (n3) -- (n4);
\end{tikzpicture}
\caption{Typical workflow of variational quantum algorithms and the underlying packages used for each task in \tcc{}. }
    \label{fig:workflow}
\end{figure}

\subsection{Conversion to qubits}
Quantum simulation of physical systems requires converting fermions (for electronic structure calculations) and bosons (for encoding nuclear wavefunctions and quantum dynamics) to qubits.
\subsubsection{Electronic Structure}
Fermionic creation and annihilation operators obey the anticommutation properties
\begin{equation}
 \begin{aligned}
    \{a_i, a^\dagger_j\} & = a_i a^\dagger_j + a^\dagger_j a_i = \delta_{ij} \\
     \{a^\dagger_i, a^\dagger_j\} & =  \{a_i, a_j\} = 0
\end{aligned}
\end{equation}
while the qubit creation operator $c^\dagger = \frac{1}{2}(X - iY)$ and annihilation operator $c =\frac{1}{2} (X+iY)$ obey the commutation properties
\begin{equation}
    \{c_i, c^\dagger_j\} = \delta_{ij} , \quad [c^\dagger_i, c^\dagger_j] =  [c_i, c_j] = 0.
\end{equation}
To comply with the required fermionic anticommutation properties,
the Jordan-Wigner transformation~\cite{JW28} can be used to map fermionic ladder operators into products of Pauli $Z$ and qubit ladder operators
\begin{equation}
    a_j = \bigotimes_{l=0}^{j-1} Z_l \otimes c_j. 
\end{equation}
Alternatively, the parity transformation~\cite{bravyi2002fermionic, seeley2012bravyi} achieves  the same goal by letting each qubit record the parity of all preceding orbitals and itself. The mapping is then
\begin{equation}
    a_j = (c_j \otimes \ket{0} \bra{0}_{j-1} - c^\dagger_j \otimes \ket{1} \bra{1}_{j-1}  ) \otimes \bigotimes_{l={j+1}}^{N-1} X_l 
\end{equation}
where $N$ is the total number of qubits.
If the ansatz conserves the total particle number, then the last qubit will not change during computation and can be removed without loss of accuracy.
If $S_z$ is also conserved, another qubit can be saved.
Thus, the parity transformation uses 2 fewer qubits than the Jordan-Wigner transformation. 
In \tcc{}, both transformations are carried out by calling \openfermion{}.

\subsubsection{Quantum Dynamics}
\label{sec:overview-conv-dynamics}
One of the key differences between quantum dynamics simulation and electronic structure calculations
is that quantum dynamics requires incorporating nuclear information in time evolution.
Similar to the case of electrons, a proper basis set must first be chosen to represent the nuclear wavefunction and operators in second-quantized form~\cite{mcardle2019digital, ollitrault2020hardware, ollitrault2020nonadiabatic}.
One popular choice is the harmonic eigenbasis, 
in which the nuclear position operator $x$ and momentum operator $p$ are expressed as
\begin{equation}
\begin{aligned}
x & = \sqrt{\frac{1}{2m\omega}} (b^\dagger + b), \ \\
p & = i\sqrt{\frac{m\omega}{2}} (b^\dagger - b), \
\end{aligned}
\end{equation}
where $m$ is the oscillator mass,
$\omega$ is the oscillator frequency and
$b^\dagger$ ($b$) are bosonic creation (annihilation) operators respectively.
The infinite boson levels are then truncated to the lowest $N$ levels for further encoding.
Other choices for the nuclear basis set include the real-space grid basis 
and a variety of discrete value representations~\cite{colbert1992novel}.
In any case, each nuclear degree of freedom is represented by a finite number of basis states $\ket{\psi} = \sum_i^N c_i \ket{i}$.

In quantum simulation, each bosonic basis vector $\ket{i}$ must be encoded into qubits, so that the system wavefunction can be described by a quantum circuit and operators can be represented by Pauli strings.
The principle for the encoding is equivalent to that of encoding integers into bit strings, and there are two common encoding strategies -- based on unary and binary encodings --
requiring $\order{N}$ and $\order{\log{N}}$ qubits, respectively~\cite{sawaya2020resource}.
Binary encoding with Gray codes is a variant of the standard binary encoding that involves changing only one bit between consecutive integers, which can be used to facilitate the efficient encoding of operators $b^\dagger$ and $b$.
\tcc{} supports all of the encoding schemes \fix{and} by default uses binary encoding with Gray codes because of its compact representation.
We note that an efficient variational encoding scheme requiring only $\order{1}$ qubits was proposed recently~\cite{li2023efficient}.

\subsection{Supported \fix{\cancel{ansatze}ans\"{a}tze}}\label{sec:supported-ansatz}
\tcc{} supports three classes of variational \fix{\cancel{ansatze}ans\"{a}tze}: (i) Unitary Coupled Cluster (UCC) -- 
 of which UCCSD, $k$-UpCCGSD, and paired UCCD (pUCCD) are variants --, (ii) Hardware-Efficient, and (iii) Variational Hamiltonian.

 \subsubsection{Unitary Coupled Cluster Ansatz}
 \label{sec:overview-ucc}
UCC \fix{\cancel{ansatze}ans\"{a}tze} have the form
 \begin{align}\label{eq:ucc}
     e^{T(\theta)}\ket{\phi}
 \end{align}
 where 
 \begin{align}\label{eq:ucc-excitations}
     T(\theta) &=\sum_{pq} \theta_{pq}\left( a^\dagger_p a_q - \text{h.c.}\right)+ \sum_{pqrs}\theta_{pqrs}\left(a^\dagger_p a^\dagger_q a_r a_s - \text{h.c.}\right)\ldots,
 \end{align}
$\theta$ denotes the vector of tunable parameters $\theta_{pq}, \theta_{pqrs}, \ldots$, and $\ket{\phi} = \ket{\text{HF}}$ is the Hartree--Fock state.  In general, excitations of any order are permitted, although it is common to only consider single and double excitations, in which case the ansatz is referred to as UCCSD.
The indices are usually confined to excitations from occupied orbitals to virtual (unoccupied) orbitals.
``Generalized'' excitations do not have to satisfy this requirement 
.
The generalized counterpart of UCCSD is UCCGSD~\cite{lee2018generalized}.

As the terms in Eq.~\ref{eq:ucc-excitations} do not necessarily commute, the exact implementation of Eq.~\ref{eq:ucc} on a gate model quantum computer is difficult.
In fact, to the best of our knowledge the ansatz defined by Eq.~\ref{eq:ucc} and Eq.~\ref{eq:ucc-excitations}
is never used in any practical VQE algorithm.
Instead, one considers a Trotterized approximation of the ansatz, also referred to as disentangled UCC~\cite{evangelista2019exact}. 
Although disentangled UCC is traditionally viewed as an approximation to the original UCC,
it can in fact exactly represent general fermionic states \fix{given appropriate operator ordering}
and should probably be considered as an alternative ansatz to the original UCC~\cite{evangelista2019exact}.
In general, the disentangled UCC ansatz can be written as
\begin{equation}
\label{eq:ucc2}
    \ket{\Psi(\theta)}_{\text{UCC}} := \prod_{k=\nex{}}^1 e^{\theta_k G_k} \ket{\phi}, \
\end{equation}
where each
$G_k$ is the anti-Hermitian excitation operator for each term in Eq.~\ref{eq:ucc-excitations}, 
and $\nex$ is  the total number of excitations. In the following, the terms $e^{\theta_k G_k}$ are referred to as UCC factors.
The UCC \fix{\cancel{ansatze}ans\"{a}tze} implemented in \tcc{} can all be viewed as special cases of Eq.~\ref{eq:ucc2}.
The specific formulation of UCCSD, $k$-UpCCGSD, and pUCCD are reviewed in the Supporting Information.

\subsubsection{Hardware-Efficient Ansatz}
Hardware-efficient \fix{\cancel{ansatze}ans\"{a}tze} are designed to be easily implementable on NISQ devices but do not necessarily preserve symmetries in molecular systems.
Finding the global minimum for such circuits can be challenging due to the vanishing gradient phenomenon and the presence of local minima~\cite{mcclean2018barren, choy2023molecular}.
However, because the required circuit depths can be much smaller than for UCC, 
hardware-efficient \fix{\cancel{ansatze}ans\"{a}tze} are widely used in quantum computational chemistry experiments performed on real quantum processors~\cite{kandala2017hardware, colless2018computation, kandala2019error, rice2021quantum, gao2021applications, kirsopp2022quantum}.
\tcc{} implements one of the most popular hardware-efficient \fix{\cancel{ansatze}ans\"{a}tze}, the $R_y$ ansatz~\cite{gao2021computational, gao2021applications,mihalikova2022cost, choy2023molecular},
whose circuit consists of interleaved layers of $R_y$ and CNOT gates
\begin{equation}
    \ket{\Psi(\theta)}_{R_y} :=  \prod_{l=k}^1 \left [ L_{R_y}^{(l)}(\theta) L_{\rm{CNOT}}^{(l)} \right ] L_{R_y}^{(0)}(\theta) \ket{\phi}, 
\end{equation}
where $k$ is the total number of layers, and the layers are defined as
\begin{equation}
\begin{aligned}
L_{\rm{CNOT}}^{(l)} & = \prod_{j=N-1}^{1} \textrm{CNOT}_{j, j+1}, \ \\
 L_{R_y}^{(l)}(\theta) & =  \prod_{j=N}^{1} \textrm{RY}_{j}(\theta_{lj}). \
\end{aligned}
\end{equation}
The gate subscripts refer to the qubit index on which the gate acts and $N$ is the total number of qubits.
An example of the ansatz is depicted in Fig.~\ref{fig:ry}.
One reason for using the $R_y$ gate is that it enforces the wavefunction coefficients to be real, which is a desired property for electronic structure problems.

As a huge number of possible hardware-efficient \fix{\cancel{ansatze}ans\"{a}tze} can be defined,
instead of providing templates for all of them, \tcc{} allows the hardware-efficient \fix{\cancel{ansatze}ans\"{a}tze} from the \qiskit{} library to be directly imported, and other customized \fix{\cancel{ansatze}ans\"{a}tze} can be defined by the user using \tc{} (see Sec.~\ref{sec:noisy-basic} for examples).

\subsubsection{Variational Hamiltonian Ansatz}
For quantum dynamics simulations, \tcc{} provides utilities to construct the 
variational Hamiltonian ansatz~\cite{wecker2015progress, miessen2021quantum}
based on user-specified Hamiltonians.
This type of ansatz \fix{depends on the Pauli strings $P_j$ contained in the Hamiltonian and thus encodes Hamiltonian information}.
Suppose the Hamiltonian is composed of $M$ Pauli strings $P_j$, as defined in Eq.~\ref{eq:ham-pauli}. The corresponding variational Hamiltonian ansatz has the form
\begin{equation}
\label{eq:dynamic-vha}
    | \Psi \rangle = \prod_l^k \prod_j^M e^{-i \theta_{lj} P_j} | \phi \rangle, 
\end{equation}
where $| \phi \rangle$ is the initial state of the system, $\theta_{kj}$ is the circuit parameter, and $k$ is a parameter defining the number of layers in the ansatz.
This heuristic construction is based on the fact that, in the short time limit, the system wavefunction $e^{-iHt}\ket{\phi}$ can be exactly described by Eq.~\ref{eq:dynamic-vha}\fix{\cancel{ with finite $k$}}.

\subsection{Engines and Backends}
\label{sec:overview-engines}
We use the term ``engine'' to refer to the methods -- such as \lstinline{"tensornetwork"} and \lstinline{"civector"} -- used to simulate the quantum circuits in \tcc{}.
The term ``backend'' refers to the underlying numerical package used to perform the simulation, such as \numpy{} and \jax{}.
The most common engine in \tcc{} is the \lstinline{"tensornetwork"} engine powered by \tc{},
which views the quantum circuit as a network of low-rank tensors and performs circuit simulation via tensor network contraction~\cite{markov2008simulating}.
This approach can offer significant advantages over full statevector simulators which may encounter memory bottlenecks.
As a result, tensor network contraction is the most frequently used method for performing large-scale quantum circuit simulations\fix{,} such as the simulation of random circuits used in quantum supremacy experiments~\cite{liu2021closing, pan2022simulation}.
\fix{
When contracting a tensor network, it is crucial to find an efficient contraction path.
The default contraction path finder in \tcc{} is based on a greedy algorithm,
which has minimal overhead, but may not be satisfactory for large quantum circuits.
The customization of contraction paths is possible via \tc{}.
}

We note that \fix{the tensor network contraction simulator} should not be confused with an MPS simulator~\cite{schollwock2011density}, although the two do share a number of things in common.
In particular, the tensor network contraction method does not assume the circuit wavefunction to adopt MPS form and does not perform any truncations to the wavefunction.

The \fix{tensor network simulator} can also be extended to the simulation of noisy circuits, using density matrix $\rho$.
\fix{The simulation of noisy quantum circuits is thus twice as memory intensive as pure state simulation.
We stress that the engine does not perform any truncation to the target density matrix, which is different from a number of preceding works~\cite{noh2020efficient, cheng2021simulating}.}
In the presence of noise, 
unfortunately, symmetries encoded in the ansatz are broken and cannot be exploited.
In \tc{}, noise channels $\mathcal{E}$ are defined by their corresponding Kraus operators $K_i$, i.e.,
\begin{equation}
    \mathcal{E}(\rho) = \sum_i K_i \rho K_i^\dagger\ .
\end{equation}
The action of $\{K_i\}$ on the circuit is achieved by transforming them into a superoperator.
\fix{
More specifically, if the channel acts on $N$ qubits, it is converted to a matrix with size $4^N\times 4^N$,
as a node in the density matrix tensor network.
We note that \tc{} supports noisy circuit simulation via Monte-Carlo sampling in addition to using density matrices, although this algorithm is not currently implemented in \tcc{}.
}

Of the various backends supported by \tcc{}, the \jax{} backend is perhaps the most versatile and can generally be used in most cases. This backend also supports GPU and TPU calculations once \jax{} is properly configured. 
If the \lstinline{"civector"} engine is used, the \numpy{} and \cupy{} backends are preferred for CPU and GPU calculations, respectively.
The reason for this is that the code in the \lstinline{"civector"} engine is highly optimized and just-in-time (JIT) compilation overheads become a bottleneck.

\subsection{Installing and contributing to \tcc{}}
\tcc{} is written in Python and can be installed via \lstinline{pip install tencirchem} 
using the command line.
The source code is hosted on the GitHub repository \url{https://github.com/tencent-quantum-lab/TenCirChem}.
We welcome all members of the quantum chemistry community to contribute to the continued development of \tcc{}.

\section{The Electronic Structure Module I. Unitary-Coupled Cluster ansatz }
\label{sec:ucc}

\subsection{Basic Usage}

The basic usage of \tcc{} can be illustrated in five lines of codes, which we demonstrate below using the UCCSD ansatz to find the minimum energy of an \ce{H_2} molecule:

\begin{lstlisting}[language=Python, caption=Simple UCCSD calculation for the hydrogen molecule., label={lst:uccsd-simple}]
from tencirchem import M, UCCSD

h2 = M(atom=[["H", 0, 0, 0], ["H", 0, 0, 0.741]])

uccsd = UCCSD(h2)
uccsd.kernel()
uccsd.print_summary(include_circuit=True)
\end{lstlisting}

The remainder of this section will explain Code Snippet~\ref{lst:uccsd-simple} in more detail.
The corresponding \fancylink{https://github.com/tencent-quantum-lab/TenCirChem/blob/master/docs/source/tutorial_jupyter/ucc_functions.ipynb}{Jupyter Notebook} is available online.

\subsubsection{Defining the molecule and basis set}
\tcc{} uses the \pyscf{} \lstinline{Mole} object to define the molecule of interest.
In other words, \lstinline{from tencirchem import M} is equivalent to \lstinline{from pyscf import M}.
The syntax for defining molecules is thus the same as that for \pyscf{}.
For example, using 3D coordinates,
atom types and coordinates  (in Angstroms) are specified in a list as shown in Code Snippet~\ref{lst:uccsd-simple}.
By default, the STO-3G minimal basis set is used, but \pyscf{} allows other basis sets to be specified via the \lstinline{basis} argument, e.g.,
\begin{lstlisting}[language=Python]
h2 = M(atom=[["H", 0, 0, 0], ["H", 0, 0, 0.741]], basis="cc-pvdz")
\end{lstlisting}

\bp{Built-in molecules.} In addition to specifying molecules by hand, the user can import a number of pre-specified molecules in STO-3G basis set from the \lstinline{tencirchem.molecule} module for debugging and fast prototyping, e.g.,
\begin{lstlisting}[language=Python]
from tencirchem.molecule import h2, h4, h8
\end{lstlisting} 
Once imported, the molecules can be used exactly as if they had been defined manually. 
The coordinates and the basis sets of these built-in molecules can be obtained by inspecting the \lstinline{Mole} object with \lstinline{m.atom} and \lstinline{m.basis} respectively.
The \lstinline{tencirchem.molecule.h2} molecule -- with a bond distance of 0.741 Å -- is frequently used in the following code examples.

\subsubsection{Specifying the ansatz and obtaining reference energies}
Once the \lstinline{h2} molecule has been specified, we initialize the UCCSD ansatz via the line
\begin{lstlisting}[language=Python]
uccsd = UCCSD(h2)
\end{lstlisting}
In addition to UCCSD, \tcc{} supports the $k$-UpCCGSD and pUCCD \fix{\cancel{ansatze}ans\"{a}tze}. These are based on the UCC base class, and can be implemented in a similar way to UCCSD once imported via:
\begin{lstlisting}[language=Python]
from tencirchem import KUPCCGSD, PUCCD
\end{lstlisting}

Ansatz information is accessible through class attributes (see Code Snippet~\ref{lst:attributes}).
The conventions for the orbital indices are described in the Supporting Information.
\begin{lstlisting}[language=Python, caption={Several useful attributes of the UCCSD class. Example corresponds to the \ce{H2} molecule.},label={lst:attributes}]
>>> # two single excitations and one double excitation
>>> uccsd.ex_ops
[(3, 2), (1, 0), (1, 3, 2, 0)]
>>> # `param_ids` maps excitation operators to parameters
>>> # some excitation operators share the same parameter due to symmetry
>>> uccsd.param_ids
[0, 0, 1]
>>> uccsd.init_guess  # generated by MP2
[0.0, -0.07260814651571333]
>>> uccsd.params # the optimized parameters
array([ 3.40632986e-17, -1.12995353e-01])
\end{lstlisting}

\bp{UCCSD class as a toolbox.} The \lstinline{UCCSD} class also allows other useful attributes to be obtained (see Code Snippet~\ref{lst:toolbox}).
The purpose is to make information in \tcc{} easy to inspect, facilitating its use as a handy toolbox.
For example, one may use the \lstinline{UCCSD} class to inspect for molecular attributes, electron integrals, reference energies and the
Jordan-Wigner transformed Hamiltonian, all of which are available 
without needing to first call the \lstinline{uccsd.kernel()} method.

\begin{lstlisting}[language=Python, caption={Several other useful attributes of the UCCSD class are also directly available without first needing to call the \lstinline{uccsd.kernel()} method.},label={lst:toolbox}]
>>> uccsd.n_qubits, uccsd.n_elec
(4, 2)
>>> # PySCF objects
>>> uccsd.mol, uccsd.hf
(<pyscf.gto.mole.Mole at 0x7f9daec92b90>, <pyscf.scf.hf.RHF at 0x7f9d92d1ec10>)
>>> # reference energies
>>> uccsd.e_hf, uccsd.e_mp2, uccsd.e_ccsd, uccsd.e_fci, uccsd.e_nuc
(-1.116706137236105,
 -1.1298675557838804,
 -1.1372745709766439,
 -1.1372744055294395,
 0.7141392859919029)
>>> # one and two-body integrals in molecular orbital basis, chemists' notation
>>> uccsd.int1e.shape, uccsd.int2e.shape
((2, 2), (2, 2, 2, 2))
>>> uccsd.h_fermion_op  # Hamiltonian as openfermion FermionOperator
0.7141392859919029 [] +
-1.2527052599711868 [0^ 0] +
-0.48227117798977825 [0^ 1^ 0 1] +
-0.6745650967143663 [0^ 2^ 0 2] +
[...... The rest of the output omitted ......]
>>> uccsd.h_qubit_op  # Hamiltonian as openfermion QubitOperator
(-0.09835117053027564+0j) [] +
(0.04531660419443148+0j) [X0 X1 X2 X3] +
(0.04531660419443148+0j) [X0 X1 Y2 Y3] +
(0.04531660419443148+0j) [Y0 Y1 X2 X3] +
[...... The rest of the output omitted ......]
\end{lstlisting}

\bp{Specifying integrals directly.} Rather than using a molecule as the input to UCCSD, one may define a UCCSD object by specifying the one and two electron integrals, without needing to explicitly refer to a molecule.  That is, applying UCCSD to the second-quantized \textit{ab initio} Hamiltonian Eq.~\ref{eq:ham-abinit}
can be done via 
\begin{lstlisting}[language=Python]
uccsd_from_integral = UCCSD.from_integral(int1e, int2e, n_elec, e_core)
\end{lstlisting}
with 
\begin{itemize}
    \item \lstinline{int1e}: an array of shape $(N,N)$  with $p,q$-th element storing the one-electron integral $h_{pq}$.
    Here $N$ is the number of spatial orbitals.
    \item \lstinline{int2e}: an array of shape $(N, N, N, N)$ with $p,q,r,s$-th element storing the two-electron integral $(pq|rs)$, in spatial orbital and chemists' notation.
    \item \lstinline{n_elec}: an integer specifying the total number of electrons in the system. Currently,  \tcc{} only supports closed-shell molecules.
    \item \lstinline{e_core}: a floating-point number for the nuclear repulsion energy $E_{\rm{nuc}}$ or the core energy if frozen occupied orbitals are involved. 
\end{itemize}

\subsubsection{Optimizing the energy}
\label{sec:ucc-optimize}
The command \lstinline{uccsd.kernel()} runs the optimization procedure to minimize the ansatz energy with respect to the variational parameters, and returns the minimum energy found. 
 By default, \tcc{} uses the L-BFGS-B optimizer implemented in \textsc{SciPy}~\cite{2020SciPy-NMeth}.
 Once \lstinline{uccsd.kernel()} has been run, the minimum ansatz  energy can be accessed by \lstinline{uccsd.energy()} or \lstinline{uccsd.e_ucc}.
The system statevector and configuration interaction vector are available by \lstinline{uccsd.statevector()} and \lstinline{uccsd.civector()}, e.g.,
\begin{lstlisting}[language=python]
>>> uccsd.civector() # configuration interaction vector
array([ 9.93623806e-01,  1.08284918e-16,  1.08284918e-16, -1.12746318e-01])
\end{lstlisting}
The optimized parameters can be obtained by \lstinline{uccsd.params}, and
the one- and two-body reduced density matrices are available 
by \lstinline{uccsd.make_rdm1()}
and \lstinline{uccsd.make_rdm2()}, respectively. 
Functions such as \lstinline{uccsd.energy}, \lstinline{uccsd.statevector}, and \lstinline{uccsd.civector} also accept custom circuit variational parameters
. For example,
\begin{lstlisting}[language=python]
>>> # if all parameters are zero, then the UCC energy is identical to HF energy
>>> uccsd.energy(np.zeros(uccsd.n_params)), uccsd.e_hf
(-1.1167061372361045, -1.116706137236105)
\end{lstlisting}

\subsubsection{Outputting quantum circuits}
\label{sec:circuit_output}

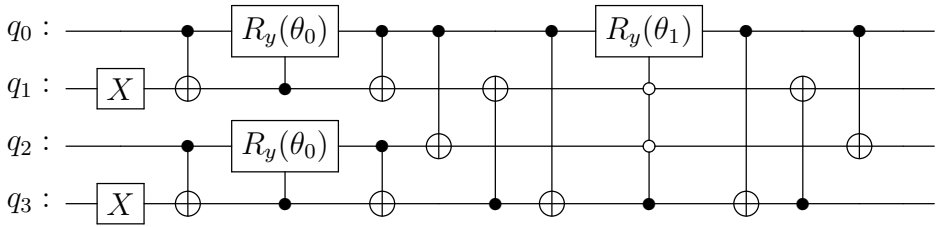
\begin{figure}[h!]
    \centering
\scalebox{1.0}{
\Qcircuit @C=1.0em @R=0.2em @!R { \\
	 	\nghost{{q}_{0} :  } & \lstick{{q}_{0} :  } & \qw & \ctrl{1} & \gate{R_y(\theta_0)} & \ctrl{1} & \ctrl{2} & \qw & \ctrl{3} & \gate{R_y(\theta_1)} & \ctrl{3} & \qw & \ctrl{2} & \qw & \qw\\
	 	\nghost{{q}_{1} :  } & \lstick{{q}_{1} :  } & \gate{X} & \targ & \ctrl{-1} & \targ & \qw & \targ & \qw & \ctrlo{-1} & \qw & \targ & \qw & \qw & \qw\\
	 	\nghost{{q}_{2} :  } & \lstick{{q}_{2} :  } & \qw & \ctrl{1} & \gate{R_y(\theta_0)} & \ctrl{1} & \targ & \qw & \qw & \ctrlo{-1} & \qw & \qw & \targ & \qw & \qw\\
	 	\nghost{{q}_{3} :  } & \lstick{{q}_{3} :  } & \gate{X} & \targ & \ctrl{-1} & \targ & \qw & \ctrl{-2} & \targ & \ctrl{-1} & \targ & \ctrl{-2} & \qw & \qw & \qw\\
\\ }}
\caption{The quantum circuit corresponding to the UCCSD ansatz for the \ce{H_2} molecule with STO-3G basis, compiled by the method \fix{\cancel{of} in Ref.~\citenum{yordanov2020efficient}}. The two $X$ gates at the left of the circuit are used to create the initial Hartree--Fock state $\ket{0101}$. The multicontrol gate applies a parameterized $R_y$ rotation to one of the four qubits, conditioned on the state of the other three.}
    \label{fig:uccsd_h2_circuit}
\end{figure}

The quantum circuit corresponding to the ansatz state can be obtained via
\begin{lstlisting}[language=Python]
c = uccsd.get_circuit()
\end{lstlisting}
which returns a \tc{} \lstinline{Circuit} object, which can then be inspected, manipulated and executed.
For example, a summary of the gates in the circuit can be obtained by
\begin{lstlisting}[language=Python]
c.gate_summary()
\end{lstlisting}
Also, conversion of the circuit from \tc{} to Qiskit format is straightforward:
\begin{lstlisting}[language=Python]
c_qiskit = c.to_qiskit()
\end{lstlisting}
By default, \tcc{} compiles circuits using the efficient method of Yordanov, Arvidsson-Shukur, and Barnes (YAB)~\cite{yordanov2020efficient, magoulas2023cnot}\fix{, } which is based on multi-qubit controlled rotations (see Fig.~\ref{fig:uccsd_h2_circuit}).
\fix{Currently triple or higher order excitations are not implemented in \tcc{}.}
\lstinline{uccsd.get_circuit(decompose_multicontrol=True)} generates the circuit with the multi-qubit controlled gate decomposed into elementary rotation gates and CNOT gates.
This decomposition is useful for noisy circuit simulation or execution on hardware. See Code Snippet~\ref{lst:noisy-ucc} for an example.
Compilation via a more traditional Trotterized method is also possible, using
\begin{lstlisting}[language=Python]
c = uccsd.get_circuit(trotter=True)
\end{lstlisting}
although this can lead to circuits significantly deeper than those compiled by the YAB method~\cite{yordanov2020efficient}. 

\subsection{Applications and benchmarking}
\label{sec:ucc-app}
\fix{
In this section, we use concrete examples to demonstrate the efficiency and flexibility of \tcc{}.
All calculations are carried out on a single computational node with the \lstinline{"civector"} engine. 
Throughout the paper, the wall time reported only accounts for VQE optimization time. Classical pre-processings, such as obtaining integrals and Hartree--Fock calculation, are not included in the reported time.
}
\subsubsection{Hydrogen chains and the hydrogen molecule}
Table\fix{\cancel{.}}~\ref{tab:ucc-hchain} gives the time required by \tcc{} to compute the minimum UCCSD energy for chains 
of increasing numbers of hydrogen atoms, using an STO-3G basis set.
Atoms in each chain are uniformly distributed with an interatomic distance set to 0.8 Å.
The largest system simulated is \ce{H16}, corresponding to 32 qubits. For such a system, storing the wavefunction in Fock space alone is already an arduous task,
while the running time for \tcc{} is still acceptable. Whereas other common quantum simulation packages usually struggle to simulate \ce{H10} (which requires 20 qubits),
\tcc{} is able to obtain the UCCSD ground state energy of this chain in 14 seconds on a single GPU node.
\fix{As a concrete example, \textsc{Qiskit-Nature} version 0.5.2 spends 34 seconds simulating \ce{H4} and 943 seconds simulating \ce{H6}, and reports an out-of-memory error when simulating larger systems.
Other packages, such as \pennylane{} and \tequilla{}, show similar performance.
We note that the high efficiency of \tcc{} is primarily due to the \lstinline{"civector"} engine and hardware acceleration plays a relatively minor role.}

\begin{table}[h]
\caption{\label{tab:ucc-hchain}
UCCSD ground state energy optimization time for hydrogen chains using \tcc{}.
CPU calculations are based on an Intel(R) Xeon(R) Platinum 8255C CPU @ 2.50GHz
while GPU calculations use the NVIDIA(R) Tesla(R) V100-PCIE-32GB. Energy errors are with respect to FCI energy, and could be reduced by adding more excitation operators in the ansatz~\cite{evangelista2019exact}.
}
\begin{tabular}{lrrrrrrr}
\hline
Molecule & Qubits & Excitations & Parameters & Device & Time & Energy  & Error \\
 &  &  &  &  & (s) & (mH)  & (mH) \\
\hline
\ce{H4 }      & 8   &  18  & 11           & CPU     & 0.05   & -2,167.55    & 0.01       \\
\ce{H6 }      & 12  &  69  & 39           & CPU     & 0.4    &  -3,204.14   & 0.27       \\
\ce{H8 }      & 16  &  200 & 108          & CPU     & 1.9    &  -4,242.52   & 0.88       \\
\ce{H10}      & 20  &  467 & 246          & GPU     & 14     & -5,282.21    & 1.37       \\
\ce{H12}      & 24  &  954 & 495          & GPU     & 51   &  -6,322.29     & 2.13       \\
\ce{H14}      & 28  &  1,749 & 899        & GPU     & 2,093  & -7,362.81    & 2.92       \\
\ce{H16}      & 32  &  2,976 & 1,520      & GPU     & 50,909  &  -8,403.63  & 3.72       \\
\hline
\end{tabular}
\end{table}

The hydrogen chain actually represents a class of molecular systems that are particularly difficult to simulate
in \tcc{}. 
In these systems, the number of spatial orbitals $N$ is the same as the number of electrons $M$.
In the following, we consider a case at the other extreme, where $M \ll N$, and simulate the potential energy curve of the $\ce{H2}$ molecule using the cc-pVTZ basis set~\cite{dunning1989gaussian}. This corresponds to $M=2$, $N=28$ and a VQE circuit on 56 qubits.
Running on a single laptop CPU, \tcc{} is able to produce all UCCSD energy values in Table~\ref{tab:ucc-h2} in 20 seconds.
Because of the small value of $M$, both the dimension of the configuration interaction space (784) and the number of excitation operators (155) are small.
\fix{
Since the \lstinline{"civector"} engine is exact, 
the small error in Table~\ref{tab:ucc-h2} is likely due to the inexactness of the disentangled UCC ansatz.
}

\begin{table}[h]
\caption{\label{tab:ucc-h2}
UCCSD/cc-pVTZ calculation of \ce{H2} in \fix{\tcc{}}, using a VQE circuit on 56 qubits. Energy errors are with respect to FCI.
}
\begin{tabular}{lrr}
\hline
Atom distance (Å) & Energy(mH)  & Error (mH) \\
\hline
0.3   & -692.85983   & 0.00000 \\
0.6   & -1153.51794   & 0.00000 \\
0.9   & -1160.41280   & 0.00001 \\
1.2   & -1112.06753   & 0.00005 \\
1.5   & -1066.16823   & 0.00014 \\
1.8   & -1034.07993   & 0.00023 \\
2.1   & -1015.56097   & 0.00028 \\
2.4   & -1006.42068   & 0.00029 \\
2.7   & -1002.39048   & 0.00028 \\
\hline
\end{tabular}
\end{table}

\subsubsection{Potential energy curve of H$_2$O}

We now compute the potential energy curve of a realistic molecule, \ce{H2O}.
The \fancylink{https://github.com/tencent-quantum-lab/TenCirChem/blob/master/example/water_pes.py}{Python script} to produce the data is available online.
This molecule has been used to benchmark many quantum algorithms~\cite{li2019quantum,ryabinkin2020iterative},
yet previous works usually use a minimal basis set or small active space.
Here, we use \tcc{} to calculate the UCCSD energy of \ce{H2O} with 6-31G(d) basis set~\cite{hehre1972self, hariharan1973influence}
with only the $1s$ orbital of the \ce{O} atom frozen.
This corresponds to 8 electrons in 17 orbitals, a quantum circuit on 34 qubits, and an ansatz involving 565 independent parameters and 1078 excitations.

\begin{figure}[htp]
    \centering
    \includegraphics[width=0.75\textwidth]{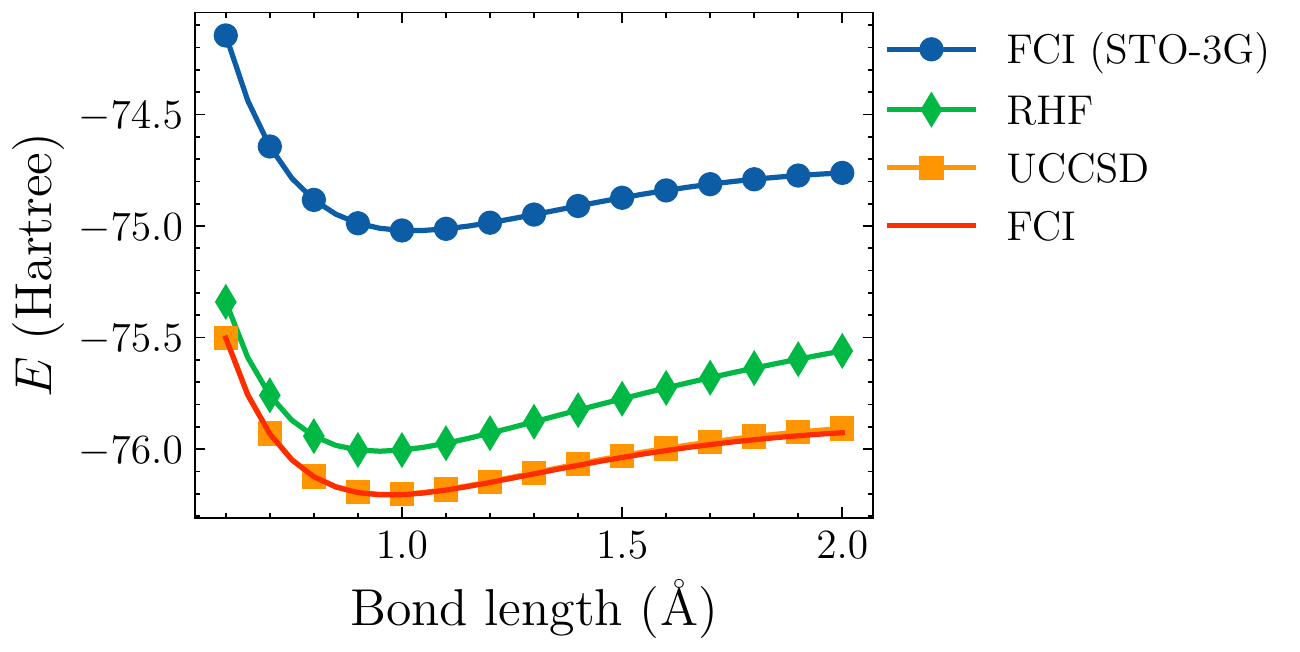}
\caption{Potential energy curve for the symmetric stretching of the O-H bond in \ce{H2O} with 6-31G(d) basis set. The $1s$ orbital of the O atom is frozen and the whole system is described by an (8e, 17o) active space and a quantum circuit on 34 qubits. The FCI energy with STO-3G basis set is included for comparison.}
    \label{fig:h2o}
\end{figure}

We study the symmetric stretching of the O-H bond with the H-O-H angle fixed at the experimental value of 104.45°~\cite{hoy1979precise}.
Results are summarized in Fig.~\ref{fig:h2o}.
The UCCSD energy by \tcc{} is very close to the FCI solution with a slight deviation observed at long bond lengths.
The equilibrium bond length is determined to be 0.97 Å, in good agreement with the experimental value (0.9584 Å).
To highlight the necessity of using an appropriate basis set,
in Fig.~\ref{fig:h2o} the FCI energy using the minimal STO-3G basis set is also given.
As expected, the energy corresponding to the STO-3G basis set is significantly higher than that corresponding to the 6-31G(d) basis set. Furthermore, the equilibrium bond distance predicted using the STO-3G basis set is 1.02 Å, 
significantly larger than the experimental value.
\fix{On average, \tcc{} takes approximately 15 minutes to produce each UCCSD energy value in Fig.~\ref{fig:h2o}, using the same GPU platform as in Table~\ref{tab:ucc-hchain}.}

\subsubsection{Hubbard model}

As a final example, we use \tcc{} and UCCSD to calculate the ground state energy of the one-dimensional Hubbard model at half-filling.
The full \fancylink{https://github.com/tencent-quantum-lab/TenCirChem/blob/master/docs/source/tutorial_jupyter/hubbard_model.ipynb}{Jupyter Notebook} is available online.
The Hamiltonian is
\begin{equation}
    H = -t \sum_{j, \sigma} (c^\dagger_{j+1, \sigma} c_{j, \sigma} + c^\dagger_{j, \sigma} c_{j+1, \sigma}) + U \sum_j n_{j \uparrow}  n_{j \downarrow}, \
\end{equation}
where $c^\dagger_{j, \sigma}$ ($c_{j, \sigma}$) creates (annihilates) an electron with spin $\sigma$ at site $j$,
$n_{j \uparrow}$ and $n_{j \downarrow}$ are electron occupation number operators at site $j$,
$t$ is the hopping integral and $U$ characterizes the Coulomb repulsion interaction strength.  We assume periodic boundary conditions.

Although the Hubbard model is not directly related to any molecular systems,
the Hamiltonian is a special case of the \textit{ab initio} Hamiltonian of Eq.~\ref{eq:ham-abinit}, and thus \tcc{} allows the construction of the corresponding \lstinline{UCCSD} objects by specifying the one-body and two-body integrals.

\begin{lstlisting}[language=Python, caption=Calculation of the UCCSD energy of half-filled Hubbard model.]
import numpy as np
from tencirchem import UCCSD

n = 6  # the number of sites
n_elec = n  # half filled
t = U = 1  # model parameters

# setup the integrals
int1e = np.zeros((n,n))
for i in range(n - 1):
    int1e[i, i + 1] = int1e[i + 1, i] = -t
int1e[n - 1, 0] = int1e[0, n - 1] = -t
int2e = np.zeros((n, n, n, n))
for i in range(n):
    int2e[i, i, i, i] = U
    
# do the calculation
uccsd = UCCSD.from_integral(int1e, int2e, n_elec)
uccsd.kernel()
\end{lstlisting}

\begin{figure}[htp]
    \centering
    \includegraphics[width=0.5\textwidth]{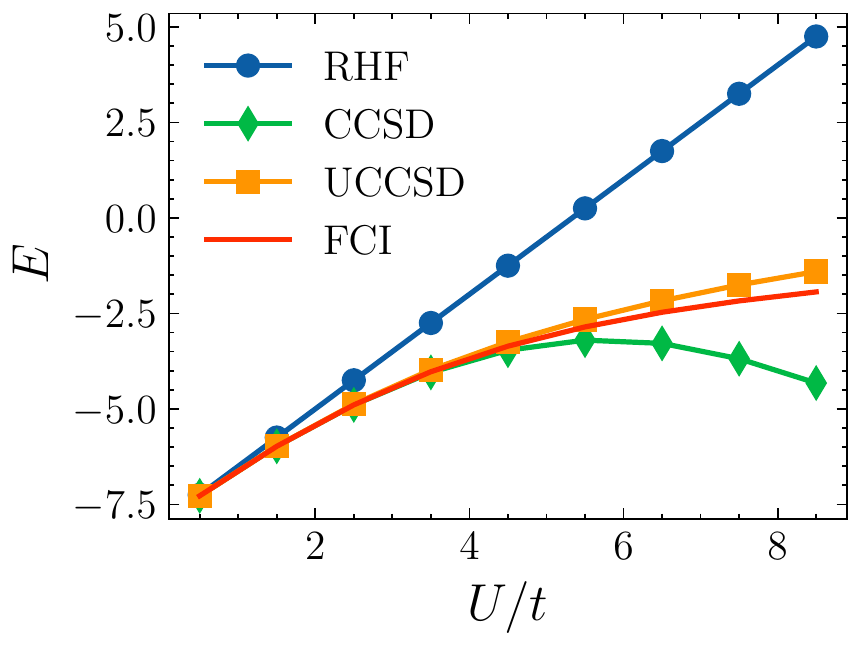}
\caption{Ground state energy of one-dimensional half-filled Hubbard model by UCCSD with increasing coulomb repulsion strength $U/t$. The UCCSD energy computed by \tcc{} tracks the FCI value significantly better than CCSD does.}
    \label{fig:hubbard}
\end{figure}

Results are shown in Fig.~\ref{fig:hubbard}.
When $U/t$ is small, both CCSD and UCCSD coincide well with the FCI solution.
As $U/t$ increases, CCSD deviates from the FCI solution, whereas the UCCSD solution tracks the FCI value much more closely, indicating that UCCSD can be better than CCSD at capturing strong correlations,
a finding consistent with \fix{a} previous report~\cite{sokolov2020quantum}.

To better recover \fix{\cancel{the}} static correlation using UCCSD, more terms can be added to the UCC ansatz
or one may consider combining UCC with other algorithms\fix{,} such as density matrix embedding theory~\cite{knizia2012density, knizia2013density, motta2017towards, mineh2022solving, li2022toward}.
\fix{On average, \tcc{} requires 0.1 seconds to calculate each UCCSD energy value in Fig.~\ref{fig:hubbard} on a regular CPU without GPU acceleration.}

\section{The Electronic Structure Module II. Hardware-efficient Ansatz and Noisy Circuit Simulation}
\label{sec:hea}
\fix{
 In real-world quantum computing systems, noise and errors are inevitable due to factors such as imperfect hardware, decoherence, and finite-shot measurement uncertainty.
Noisy circuit simulators allow researchers to simulate the effects of noise on quantum circuits and get a better understanding of the behavior of quantum algorithms in realistic conditions.
}
In this section, we use the hydrogen molecule $\rm{H}_2$ to illustrate
how to perform noisy circuit simulation in \tcc{}. After using the parity transformation, the corresponding Hamiltonian acts on 2 qubits.
The \fancylink{https://github.com/tencent-quantum-lab/TenCirChem/blob/master/docs/source/tutorial_jupyter/noisy_simulation.ipynb}{Jupyter Notebook} for this section is available online.

\subsection{Basis Usage}
\label{sec:noisy-basic}
\tcc{} uses the \lstinline{HEA} (Hardware-Efficient Ansatz) class for noisy circuit simulation, 
which has a different interface from the UCC classes encountered in Sec.~\ref{sec:ucc}. 
A primary motivation for using HEA \fix{\cancel{ansatze}ans\"{a}tze} is that almost all UCC circuits (with perhaps pUCCD circuits being the exception) are too large for both noisy circuit simulation as well as execution on currently available quantum devices.  In contrast, HEA circuits are, by design, constructed to be implementable on whatever device is available.

\fix{
Another reason is that, from a user's perspective, the \lstinline{HEA} class ought to provide a greater degree of low-level customization compared to the UCC classes.
For example, for noiseless UCC simulation, the ansatz is defined by excitation operators and compilation to native gates is of less interest. Conversely, in the context of noisy HEA circuit simulation, it is imperative to give users complete control over the ansatz circuit.
}
Nevertheless, noisy UCC circuit simulation can be carried out within \tcc{} without much additional effort, which will be demonstrated in Sec.~\ref{sec:noisy-basic-ansatz}.

\subsubsection{Specifying the Hamiltonian and the ansatz}
\label{sec:noisy-basis-inputs}
Unlike \lstinline{UCC}, the \lstinline{HEA} class 
is not initialized by specifying \fix{\cancel{molecular inputs}the molecule}.
Rather, \lstinline{HEA} takes as inputs (i) the Hamiltonian in \lstinline{openfermion.QubitOperator} form 
and (ii) the circuit ansatz.

Within \openfermion{}, the Hamiltonian can be constructed by interfacing with quantum chemistry packages\fix{,}  such as \pyscf{} using
\lstinline{openfermion.MolecularData}.
\begin{lstlisting}[language=python, caption={Construction of the system Hamiltonian in fermion operator form using \openfermion.}]
from openfermion import MolecularData
from openfermionpyscf import run_pyscf

geometry = [["H", [0, 0, 0]], ['H', [0, 0, 0.741]]]
basis = "STO-3G"
multiplicity = 1
molecule = MolecularData(geometry, basis, multiplicity)
molecule = run_pyscf(molecule)
h_fermion_op = molecule.get_molecular_hamiltonian()
\end{lstlisting}

An alternative approach is to use the \lstinline{UCC} class attributes\fix{,}  such as \lstinline{uccsd.h_fermion_op}. 

\begin{lstlisting}[language=Python, caption={Construction of the system Hamiltonian in qubit operator form using \lstinline{UCCSD} class attributes.}, label={lst:ham-hea}]
from tencirchem import UCCSD, parity
from tencirchem.molecule import h2

uccsd = UCCSD(h2)
# Hamiltonian as openfermion.FermionOperator
h_fermion_op = uccsd.h_fermion_op
# use parity transformation for qubit Hamiltonian and remove 2 qubits
h_qubit_op = parity(h_fermion_op, uccsd.n_qubits, uccsd.n_elec)
\end{lstlisting}

\begin{figure}

    \centering
\scalebox{1.0}{
\Qcircuit @C=1.0em @R=1.0em @!R { \\
\nghost{{q}_{0} :  } & \lstick{{q}_{0} :  } & \gate{R_y(\theta_0)} &\ctrl{1} & \gate{R_y(\theta_2)} & \qw \\
\nghost{{q}_{1} :  } & \lstick{{q}_{1} :  } & \gate{R_y(\theta_1)}  & \targ & \gate{R_y(\theta_3)} & \qw \\
\\ }}
\caption{Circuit diagram for a 2-qubit 1-layered $R_y$ ansatz. More layers can be added by repeating the CNOT and $R_y$ rotation unit.}
\label{fig:ry}
\end{figure}
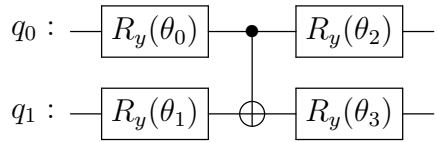

The ansatz is defined using \tc{}.
In the example below, we construct a 1-layer $R_y$ ansatz comprising 4 parameters
and the circuit diagram is depicted in Fig.~\ref{fig:ry}.
The initial parameters create the HF state $\ket{01}$  under parity transformation.
\fix{
In addition to defining the ansatz from scratch,
it is also possible to use a pre-defined ansatz, which will be discussed more thoroughly in Sec.~\ref{sec:noisy-basic-ansatz}.
}

\begin{lstlisting}[language=Python, caption=Defining the 1-layer $R_y$ ansatz using \tc{}.]
import numpy as np
from tensorcircuit import Circuit

n_qubits = 2
n_params = 4
init_guess = [0, np.pi, 0, 0]
def get_circuit(params):
    c = Circuit(n_qubits)
    # the ansatz body
    c.ry(0, theta=params[0])
    c.ry(1, theta=params[1])
    c.cnot(0, 1)
    c.ry(0, theta=params[2])
    c.ry(1, theta=params[3])
    return c
\end{lstlisting}

\subsubsection{Noisy circuit simulation and optimization}
The remaining workflow is similar to the \lstinline{UCCSD} case: (i) create an \lstinline{HEA} instance, (ii) 
run the kernel, and (iii) print the summary:
\begin{lstlisting}[language=Python]
from tencirchem import HEA
hea = HEA(h_qubit_op, get_circuit, init_guess, engine="tensornetwork-noise")
hea.kernel()
hea.print_summary()
\end{lstlisting}
The default optimizer used is again L-BFGS-B from \scipy{}.

The gradients are evaluated by the parameter-shift rule~\cite{mitarai2018quantum, schuld2019evaluating}.
The parameter-shift rule uses two separate circuits to evaluate the gradient of one parameter.
Its essence can be summarized using a single-qubit single-gate illustrative example.
Suppose the circuit wavefunction is $\ket{\psi} = e^{\theta R} \ket{\phi}$ with $R^2 = -I$.
This leads to $\ket{\psi} = (\cos{\theta} + \sin{\theta} R ) \ket{\phi}$, and the energy expectation is then
\begin{equation}
    \braket{E}(\theta) = \cos^2{\theta} \braket{\phi|H|\phi} + \sin{\theta}\cos{\theta} \braket{\phi|RH + HR|\phi}
     + \sin^2{\theta} \braket{\phi|RHR|\phi}\ .
\end{equation}
Thus
\begin{equation}
    \pdv{\braket{E}}{\theta} = \braket{E}(\theta + \frac{\pi}{4}) - \braket{E}(\theta - \frac{\pi}{4})\ .
\end{equation}
The expression is slightly different from the literature where $e^{\frac{\theta}{2}R}$ is usually assumed.

By default, using the \lstinline{"tensornetwork-noise"} engine (line 2) adds isotropic depolarizing noise\fix{~\cite{nielsen2002quantum}} with error probability $p=0.02$
\begin{equation}
    \mathcal{E}(\rho) = (1 - p)\rho + \frac{p}{15} \sum_{i=1}^{15} \mathcal{P}_j \rho \mathcal{P}_j. \
\end{equation}
to each CNOT gate in the circuit. Here $\rho$ is the density matrix, and the $\mathcal{P}_j$ are the two-qubit Pauli matrices (excluding the identity operator $I$).
 Hereafter we will also refer to $p$ as the CNOT error probability. 
 The value $p=0.02$ corresponds to approximately 98\% CNOT average gate fidelity, defined as
 \begin{equation}
     F = \int \braket{\psi|\mathcal{E}(\ket{\psi}\bra{\psi})|\psi} d \psi = 1 - \frac{4}{5}p. \
 \end{equation}
Single qubit gate noise is not included. The energy obtained in this case is -1.1203.

Other noise models can be specified using the \lstinline{NoiseConf} class from \tc{}, as illustrated below:

\begin{lstlisting}[language=Python, caption={A simple customized noise model with depolarizing error probability $p=0.25$, corresponding to average gate fidelity of  80\%. 
The energy obtained is -0.9245, higher than the value of -1.1203 obtained from the $p=0.02$ case.}]
from tensorcircuit.noisemodel import NoiseConf
from tensorcircuit.channels import isotropicdepolarizingchannel

engine_conf = NoiseConf()
# larger noise, corresponding to 80%
channel = isotropicdepolarizingchannel(p=0.25, num_qubits=2)
engine_conf.add_noise("cnot", channel)

hea = HEA(h_qubit_op, get_circuit, init_guess, \
          engine="tensornetwork-noise", engine_conf=engine_conf)
hea.kernel()
hea.print_summary()
\end{lstlisting}

In addition to the depolarizing channel, \tc{} supports the amplitude damping channel, phase damping channel, and thermal relaxation channel\fix{~\cite{nielsen2002quantum}}.

The \lstinline{"tensornetwork-noise"} engine does not consider measurement uncertainty. Measurement uncertainty can be accounted for using the \lstinline{"tensornetwork-noise&shot"} engine, as shown below:

\begin{lstlisting}[language=Python, caption={Accounting for measurement uncerainty in noisy circuit simulation using the \lstinline{"tensornetwork-noise&shot"} engine.}]
>>> # number of shots: 4096
>>> for i in range(5):
>>>    print(hea.energy(engine="tensornetwork-noise&shot"))
-0.9209869918508058
-0.9306141224500795
-0.9176478740443613
-0.929275708950061
-0.9241459510180587
>>> # number of shots: 4096 * 128"
>>> hea.shots = 4096 * 128
>>> for i in range(5):
>>>    print(hea.energy(engine="tensornetwork-noise&shot"))
-0.9246797253478469
-0.9251027083245665
-0.9243841547125008
-0.9236478492038139
-0.9248303187894168
\end{lstlisting}
By default, the number of measurement shots made for each term in the Hamiltonian is 4096.  Increasing the number of measurement shots decreases the energy uncertainty.
Here the engine is switched temporarily at runtime,
and specifying the engine while initializing the \lstinline{HEA} class is also viable, 
by \lstinline{hea=HEA(*args, engine=engine)}. 

\fix{
If desired, a real quantum hardware engine can be specified by setting the engine to \lstinline{"qpu"}.
The circuit is then executed on a 9-qubit super-conducting QPU hosted by the Tencent Quantum Lab quantum cloud service.
A private token is required for successful execution.
The platform is currently under closed beta and we are developing more features. 
The access token and configuration documents can be requested by sending an email to the authors.
}

If noiseless results are desired for the HEA ansatz, this can be achieved by using the \lstinline{"tensornetwork"} engine, i.e.,  \lstinline{hea.energy(engine="tensornetwork")}.  In this case, the energy is in exact agreement with the FCI energy for the hydrogen molecule system.

\subsubsection{Using pre-defined \fix{\cancel{ansatze}ans\"{a}tze}}
\label{sec:noisy-basic-ansatz}
At the end of the section, we show how to use pre-defined ansatze for noisy circuit simulation.
TenCirChem has implemented the $R_y$ ansatz.
In the following code the \lstinline{HEA} instance is rebuilt using the \lstinline{HEA.ry} function, 
which constructs the qubit Hamiltonian and the $R_y$ ansatz automatically.
The result after running the kernel is exactly the same as what we've illustrated step by step above.

\begin{lstlisting}[language=Python]
hea = HEA.ry(uccsd.int1e, uccsd.int2e, uccsd.n_elec, uccsd.e_core,  \
             n_layers=1, engine="tensornetwork-noise")
hea.kernel()
\end{lstlisting}

If desired, the UCC ansatz defined in the \lstinline{UCC} class can be fed into the \lstinline{HEA} class for noisy circuit simulation. 
The following code snippet shows how to simulate the UCCSD circuit of the \ce{H2} molecule in the presence of noise.
As described in Sec.~\ref{sec:circuit_output}, \tcc{} by default uses a multi-qubit controlled $R_y$ gate for UCC circuit simulation, which has to be decomposed into elementary gates before actual execution on quantum computers. 
The gradient is turned off because the parameter-shift rule is not directly applicable to the circuit
in which multiple gates share the same parameter.
The COBYLA optimizer implemented in \scipy{}~\cite{2020SciPy-NMeth} is used for optimization.
The energy obtained is -0.8358, which is significantly higher than \fix{the energy by the} HEA ansatz, because there are 18 noisy CNOT gates in the circuit.
\fix{We comment that simulating UCC circuits with the \lstinline{HEA} class is non-standard.
Usually, it is necessary to simplify the ansatz before execution under realistic hardware conditions~\cite{peruzzo2014variational, o2016scalable}.}

\begin{lstlisting}[language=Python, caption={Noisy simulation of UCC circuits.}, label={lst:noisy-ucc}]
from functools import partial
get_circuit = partial(uccsd.get_circuit, decompose_multicontrol=True)
hea = HEA(uccsd.h_qubit_op, get_circuit, uccsd.init_guess, \
          engine="tensornetwork-noise")
hea.grad = "free"
hea.kernel()
\end{lstlisting}

\subsection{Applications}

\subsubsection{Effect of gate noise on VQE energy}
\label{sec:app-noise}
\begin{figure}[htp]
    \centering
    \includegraphics[width=0.7\textwidth]{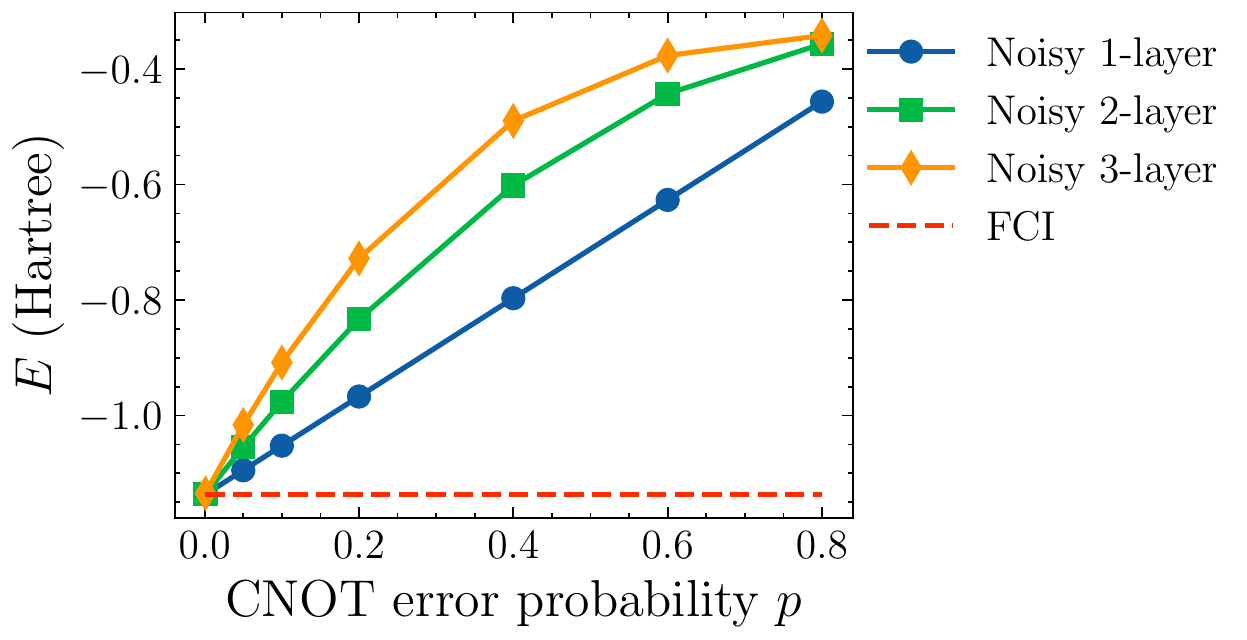}
\caption{The VQE ground state energy of \ce{H2} with parity transformation in the presence of depolarizing error, characterized by the error probability $p$. The $R_y$ hardware-efficient ansatz is used, with between 1 and 3 layers.}
    \label{fig:error-probability}
\end{figure}

As a first application, we show how the CNOT depolarizing error affects the optimized VQE energy. Results are shown in Fig.~\ref{fig:error-probability}.
The $R_y$ hardware-efficient ansatz and the \ce{H2} molecule with parity transformation are used, as in Sec.~\ref{sec:noisy-basic}.

We additionally test how the number of ansatz layers affects the obtained energy.
If a one-layer ansatz is adopted, the energy increases linearly with the error probability up to $p=0.8$. This is a consequence of the fact that this ansatz, when compiled into a quantum circuit, contains only a single CNOT gate.
Increasing the number of layers does not lead to more accurate energy estimation. 
Rather, the noise caused by more CNOT gates in the ansatz further degrades the results.
\tcc{} requires approximately 1 second to produce each data point in Fig.~\ref{fig:error-probability}, which involves iterating the VQE parameter optimization procedure until convergence.

\subsubsection{Effect of measurement shots on VQE energy uncertainty}
\begin{figure}[htp]
    \centering
    \includegraphics[width=0.5\textwidth]{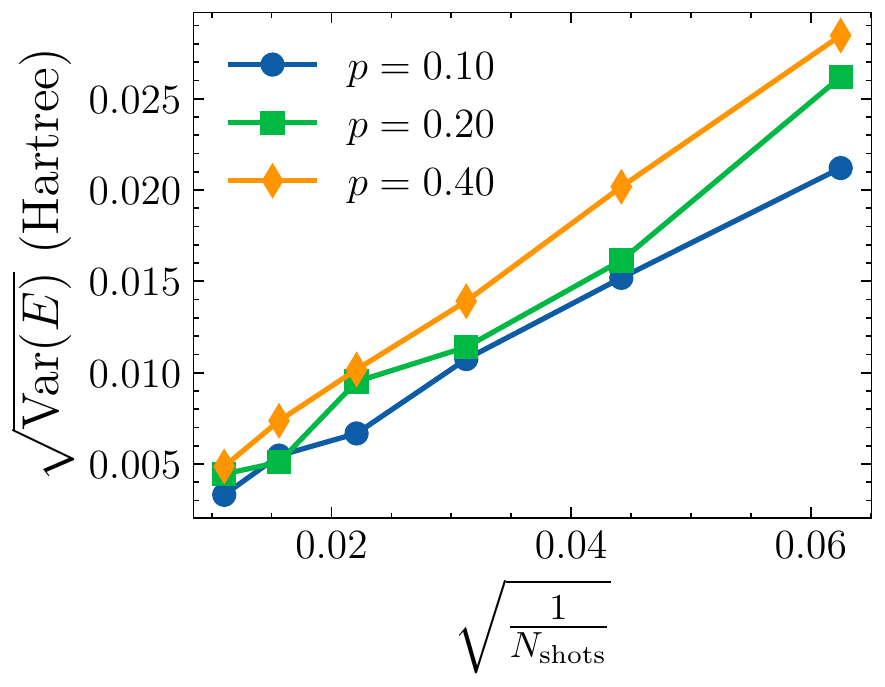}
\caption{The VQE energy in the presence of depolarizing error, characterized by the error probability $p$. The $R_y$ circuit with 1 layer is used as the ansatz.}
    \label{fig:uncertainty}
\end{figure}

Here we show how the number of measurement shots $\nshots$ affects the standard deviation of the optimized VQE energy. Results are shown in Fig~\ref{fig:uncertainty}.
The \ce{H2} molecule of Sec.~\ref{sec:app-noise} is used with a single layer $R_y$ ansatz.
We test $\nshots$ values from $2^8$ to $2^{13}$.
For each value of $\nshots$ investigated, the system energy $E$ is evaluated 64 times and the standard deviation is calculated.
Note that the circuit parameters are kept constant at the optimal value and the VQE iteration is not run.

As expected, the standard deviation of $E$ is proportional to $\sqrt{\frac{1}{\nshots}}$~\cite{mcclean2016theory}.
Increasing the error probability $p$ causes a larger standard deviation of the energy, although the effect is far less significant than that caused by adjusting $\nshots$.
\tcc{} requires approximately 1 second to produce a single data point in Fig.~\ref{fig:uncertainty}, each of which involves 64 separate energy evaluations.

\begin{figure}[htp]
    \centering
    \includegraphics[width=0.5\textwidth]{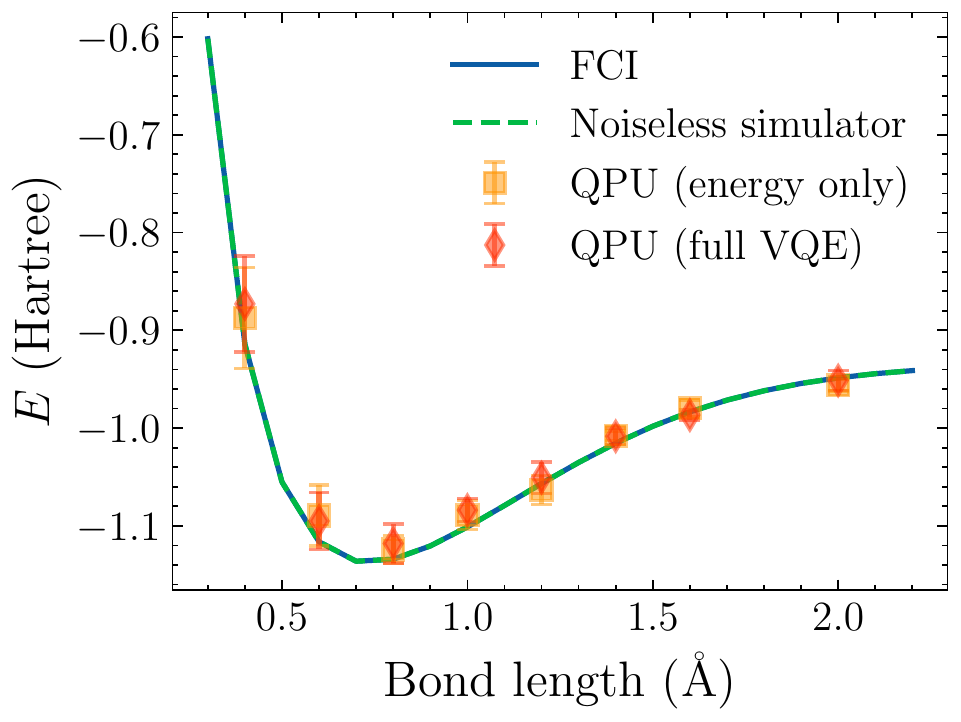}
\caption{\fix{The VQE potential energy curve of \ce{H2} computed on a QPU backend. QPU energies obtained from the classically optimized parameter and full VQE optimization process are both shown.}}
    \label{fig:h2_qpu}
\end{figure}

\fix{
\subsubsection{Running quantum hardware experiments}
In this section, we use QPUs to compute the VQE energy of the \ce{H2} molecule, and the results are summarized in Fig.~\ref{fig:h2_qpu}.
We use the same one-layered $R_y$ ansatz described in previous sections.
To reduce the computational cost, the parameter vector is reduced to $[\theta, \pi, 0, 0]$ with only one variable $\theta$.
Noiseless simulation shows that using this ansatz is sufficient to reach the FCI energy.
On QPU, 8192 shots are taken to determine the expectation value of each term in the Hamiltonian.
The error is mitigated by standard readout error mitigation.
For each set of circuit parameters, the energy is evaluated 8 times to produce the final energy expectation and its standard deviation.
We first compute the potential energy curve using the classically optimized parameter and then run the full VQE optimization starting from $\theta=0$.
In both cases, the QPU engine can correctly describe the dissociation curve of the \ce{H2} molecule.
Moreover, we observe that the energy standard deviation increases as the bond length decreases.
This is because at short bond lengths the Pauli strings have large coefficients due to strong interaction between the particles.
}

\section{The Quantum Dynamics Module}
\label{sec:dynamics}

\subsection{Basic Usage}

\subsubsection{Defining the Hamiltonian and basis set}
The first step in quantum dynamics simulations is specifying the Hamiltonian.
In this section, we focus on the spin-boson model with 1 bosonic mode:
\begin{equation}
\label{eq:dynamic-sbm}
H = \frac{\epsilon}{2} \sigma_z + \Delta \sigma_x + \omega  b^\dagger b + g \sigma_z ( b^\dagger +  b)    
\end{equation}
where $\sigma_z = Z$ and $\sigma_x = X$ are Pauli matrices, $\epsilon$ is the eigenfrequency and $\Delta$ is the tunnelling rate.
\fix{
The spin-boson model is a classical model for a variety of chemical processes, 
such as electron transfer and photochemistry~\cite{leggett1987dynamics}.
}
For ease of demonstration, we truncate the allowed boson states to two levels. 
More sophisticated examples can be found in Sec.~\ref{sec:dynamic-app} and in the online \fancylink{https://github.com/tencent-quantum-lab/TenCirChem/blob/master/docs/source/tutorial_jupyter/sbm_dynamics.ipynb}{Jupyter Notebook}.

Defining the system Hamiltonian and basis sets is performed as follows:

\begin{lstlisting}[language=Python, caption=Defining the system Hamiltonian and basis set for 1-mode spin-boson model.]
from tencirchem import Op, BasisHalfSpin, BasisSHO

epsilon = 0
delta = 1
omega = 1
g = 0.5

ham_terms = [
    Op("sigma_z", "spin", epsilon),
    Op("sigma_x", "spin", delta),
    Op(r"b^\dagger b", "boson", omega),
    Op("sigma_z", "spin", g) * Op(r"b^\dagger+b", "boson")
]
basis = [BasisHalfSpin("spin"), BasisSHO("boson", omega=omega, nbas=2)]
\end{lstlisting}
The \lstinline{Op} and the basis set classes are directly imported from the \reno{} package.
Each term in the Hamiltonian takes three arguments: (i) the operator symbol; (ii) the name of the associated degree of freedom; and (iii) the (optional) coefficient.

In setting the basis sets (line 14), \lstinline{BasisSHO} refers to the simple harmonic oscillator basis,
and \lstinline{nbas=2} restricts the phonon states to the lowest two levels.

Conversion of the Hamiltonian and basis sets into qubits is then performed in \tcc{} using the \lstinline{qubit_encode_op} and \lstinline{qubit_encode_basis} commands, as illustrated below:

\begin{lstlisting}[language=Python, caption={Transformation from physical basis set into qubit spin basis.  Here we use the binary encoding with Gray code approach (see Sec.~\ref{sec:overview-conv-dynamics}) to encode the bosonic states.}]
from tencirchem.dynamic import qubit_encode_op, qubit_encode_basis
boson_encoding = "gray"
ham_terms_spin, constant = qubit_encode_op(ham_terms, basis, boson_encoding)
basis_spin = qubit_encode_basis(basis, boson_encoding)
\end{lstlisting}

After transformation, the Hamiltonian and basis sets can be inspected as follows:
\begin{lstlisting}[language=Python, caption=Inspection of the transformed Hamiltonian.]
>>> ham_terms_spin
[Op('X', ['spin'], 1.0),
 Op('Z', [('boson', 'TCCQUBIT-0')], -0.5),
 Op('Z X', ['spin', ('boson', 'TCCQUBIT-0')], 0.5)]
>>> basis_spin
[BasisHalfSpin(dof: spin, nbas: 2),
 BasisHalfSpin(dof: ('boson', 'TCCQUBIT-0'), nbas: 2)]
\end{lstlisting}
As expected, the Hamiltonian is now expressed in terms of Pauli strings and the basis sets are all transformed to the spin-$\frac{1}{2}$ basis set.
\fix{
Here, the term in square brackets in each operator of the above Code Snippet represents the label(s) of the spin-$\frac{1}{2}$ basis on which the Pauli operators are defined.
}
As we restrict to two phonon levels, only one spin-$\frac{1}{2}$ basis set \fix{with label \lstinline{('boson', 'TCCQUBIT-0')}} is generated, i.e., the phonon mode is represented by a single qubit.

\subsubsection{Construct the ansastz}

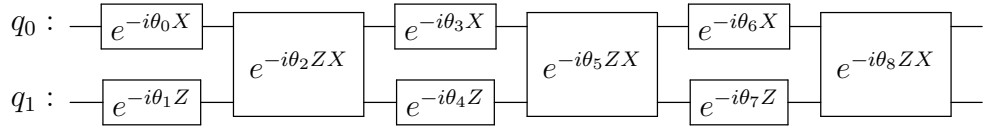
\begin{figure}[h]
    \centering
\scalebox{1.0}{
\Qcircuit @C=1.0em @R=1.0em @!R { \\
\nghost{{q}_{0} :  } & \lstick{{q}_{0} :  } & \gate{e^{-i\theta_0 X}} & \multigate{1}{e^{-i\theta_2ZX}} & \gate{e^{-i\theta_3 X}} & \multigate{1}{e^{-i\theta_5ZX}} & \gate{e^{-i\theta_6 X}} &  \multigate{1}{e^{-i\theta_8ZX}} & \qw \\
\nghost{{q}_{1} :  } & \lstick{{q}_{1} :  } & \gate{e^{-i\theta_1 Z}} &  \ghost{e^{-i\theta_2ZX}}  & \gate{e^{-i\theta_4 Z}} & \ghost{e^{-i\theta_5ZX}} & \gate{e^{-i\theta_7 Z}} & \ghost{e^{-i\theta_8ZX}}& \qw \\
\\ }}
\caption{Quantum circuit corresponding to the 3-layered variational Hamiltonian ansatz for the 1-mode spin-boson model using binary encoding with Gray code. $q_0$ represents the spin and $q_1$ represents  the boson.}
    \label{fig:dynamics_circuit}
\end{figure}

The variational Hamiltonian ansatz (see Fig.~\ref{fig:dynamics_circuit}) defined in Eq.~\ref{eq:dynamic-vha} can now be constructed using the \lstinline{get_ansatz} function, and the Jacobian then obtained by \lstinline{get_jacobian_func}:

\begin{lstlisting}[language=Python, caption=Construction of the ansatz and the function to evaluate the Jacobian.]
import tensorcircuit as tc
from tencirchem import set_backend
from tencirchem.dynamic import get_ansatz, get_jacobian_func

# dynamics simulation requires auto-differentiation from JAX.
set_backend("jax")

# the initial state
init_circuit = tc.Circuit(len(basis_spin))
# number of layers 
n_layers = 3
# get the ansatz. Note that the spin basis is fed in
ansatz = get_ansatz(ham_terms_spin, basis_spin, n_layers, init_circuit)

# ansatz accepts parameters and outputs wavefunction
import numpy as np
print(ansatz(np.zeros(n_layers * len(ham_terms_spin))))
# the function to evaluate the jacobian of the wavefunction
jacobian_func = get_jacobian_func(ansatz)
\end{lstlisting}

Here, \lstinline{basis_spin} defines the ordering of the qubits in the circuit. 
More specifically, the first and second elements in \lstinline{basis_spin} refer to the spin and boson respectively, and thus in the quantum circuit the first qubit represents the spin and the second qubit represents the boson. Note that the backend needs to be set to \jax{} explicitly because the default \numpy{} backend does not support automatic differentiation.

\subsubsection{Time evolution}
With the ansatz function to calculate the wavefunction and its Jacobian, 
the equation of motion Eq.~\ref{eq:dynamic-mv} can now be readily solved to determine $\dot \theta_k$.
The following code wraps the derivative function in \scipy{} format and solves the initial value problem. 
The wavefunction is evolved from time $t=0$ to $t=10$ using a time interval of $0.1$.

\begin{lstlisting}[language=Python, caption=Time evolution of the spin-boson system.]
from tencirchem import get_dense_operator
from tencirchem.dynamic import get_deriv
# the Hamiltonian in dense matrix format
h = get_dense_operator(basis_spin, ham_terms_spin)
# time derivative for circuit parameters in the scipy solve_ivp format
def scipy_deriv(t, _theta):
    return get_deriv(ansatz, jacobian_func, _theta, h)
from scipy.integrate import solve_ivp
# time step
tau = 0.1
# initial value
theta = np.zeros(n_layers * len(ham_terms_spin))
for n in range(100):
    # time evolution
    scipy_sol = solve_ivp(scipy_deriv, [n * tau, (n + 1) * tau], theta)
    # time evolved parameter
    theta = scipy_sol.y[:, -1]
\end{lstlisting}

Note that \tcc{} also provides a high-level interface for simulating dynamics, e.g., 

\begin{lstlisting}[language=Python, caption=Time evolution of the spin-boson system using the \lstinline{TimeEvolution} class.]
from tencirchem import TimeEvolution

te = TimeEvolution(ham_terms, basis)
for i in range(100):
    te.kernel(0.1)
\end{lstlisting}

Using this interface, the user need only specify the Hamiltonian and the basis sets, 
and encoding to qubits is carried out automatically using binary encoding with Gray code.
\fix{Configuration of the encoding strategy, the circuit initial condition, and the number of layers in the ansatz are also supported.}

\subsection{Implementation}

\tcc{} uses classical matrix manipulations to calculate the $M$ matrix and $V$ vector in the equation of motion (Eq.~\ref{eq:eom-theta}) instead of 
faithfully simulating the quantum circuit.
The first step is to obtain the Jacobian $\pdv{\ket{\Psi}}{\theta_k}$ by forward-mode auto-differentiation via the \jax{} backend.
\fix{
Forward-mode auto-differentiation is preferred over reverse-mode auto-differentiation in this case, because usually there are more amplitudes in $\ket{\Psi}$ than parameters $\theta_k$.
The whole matrix $M$ is then calculated by a single matrix-matrix multiplication according to Eq.~\ref{eq:dynamic-mv}.
Similarly, the whole vector $V$ is calculated by computing $H\ket{\psi}$ via a single matrix-vector multiplication, and then another matrix-vector multiplication between the Jacobian and $H\ket{\psi}$.
}
This implementation minimizes the just-in-time compile time in \jax{},
because evaluating the Jacobian is the only bottleneck that needs to be compiled during the workflow.
The $M$ matrix is usually not invertible due to linear dependencies in $\pdv{\ket{\Psi}}{\theta_k}$.
It\fix{\cancel{'s}s} eigenvalues $\lambda_i$ are modified with $\lambda_i \rightarrow \lambda_i + \epsilon e^{-\lambda_i / \epsilon}$
for regularization, where $\epsilon$ is a pre-defined small number, typically $1 \times 10^{-5}$.

\subsection{Applications}
\label{sec:dynamic-app}

\subsubsection{Spin relaxation of spin-boson model}
\begin{figure}[htp]
    \centering
    \includegraphics[width=0.7\textwidth]{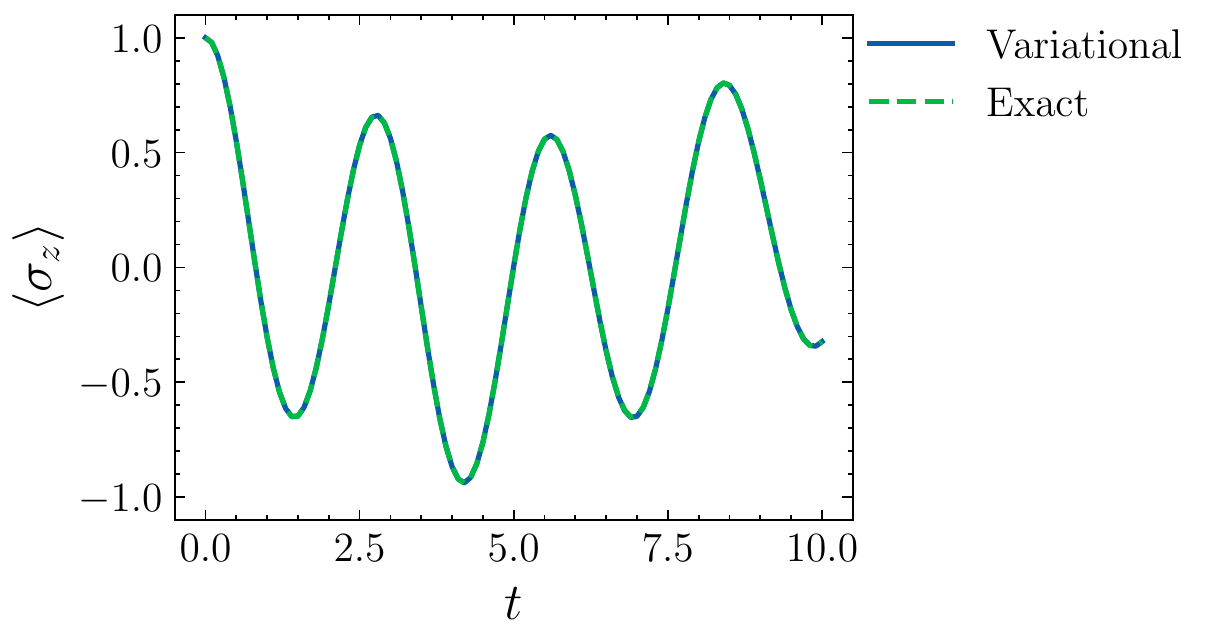}
\caption{Spin relaxation of the 1-mode spin-boson model simulated by variational quantum dynamics.}
    \label{fig:sbm}
\end{figure}

Here we show the simulated dynamics of the 1-mode spin-boson model defined in Eq.~\ref{eq:dynamic-sbm}, with parameters $\epsilon = 0$, $\Delta = 1$, $\omega = 1$ and $g = 0.5$.
Unlike the previous example which truncated the boson to the lowest two states, here we allow 8 bosonic states, and the corresponding circuit acts on 4 qubits.
The simulated results are shown in Fig.~\ref{fig:sbm}, and are in good agreement with the solution with exact diagonalization.

\subsubsection{Marcus inverted region of the charge transfer model}
In this section, we simulate the charge transfer or oxidation-reduction reaction between two molecules using the Marcus model~\cite{shuai2020applying}
\begin{equation}
    H = -V(a^\dagger_0 a_1 + a^\dagger_1 a_0) + \Delta G a^\dagger_1 a_1 + \omega \sum_{i=0, 1} b^\dagger_i b_i + g \omega \sum_{i=0, 1} a^\dagger_i a_i (b^\dagger_i + b_i). \
\end{equation}
Here $a_0$ and $a_1$ are annihilation operators for the charge on the first and the second molecule respectively.
We assume transfer integral $V=-0.1$, dimensionless coupling constant $g=1$, and vibration frequency $\omega=0.5$.
The Marcus charge transfer theory predicts that, by decreasing the reaction Gibbs free energy change $\Delta G$, the reaction rate $k$ will first increase and then decrease, according to
\begin{equation}
\label{eq:marcus}
 k = \frac{V^2}{\hbar} \sqrt{\frac{\pi\beta}{\lambda}} \exp{-\frac{\beta(\lambda + \Delta G)^2}{4\lambda}}   
\end{equation}
where $\lambda = 2g^2\omega$ is the reorganization energy and $\beta$ is the inverse temperature.
Due to the quadratic term on the exponential, the maximum rate is reached when $\Delta G=-\lambda=-1$.

In the simulation,
we again truncate the number of boson states to 8, and thus the Hamiltonian is encoded into 7 qubits using the Gray code method.
The charge is initially set to be located at the first molecule and the nuclear coordinates are relaxed
according to the localized charge~\cite{kloss2019multiset}.
The system is then evolved to $t=8$ using \tcc{} and we compute $\braket{a^\dagger_0 a_0}$ (the charge occupation on the first molecule) as a function of time, with $\Delta G$ ranging from 0 to -2.
The full \fancylink{https://github.com/tencent-quantum-lab/TenCirChem/blob/master/docs/source/tutorial_jupyter/marcus.ipynb}{Jupyter Notebook} is available online.

\begin{figure}[htp]
    \centering
    \includegraphics[width=0.8\textwidth]{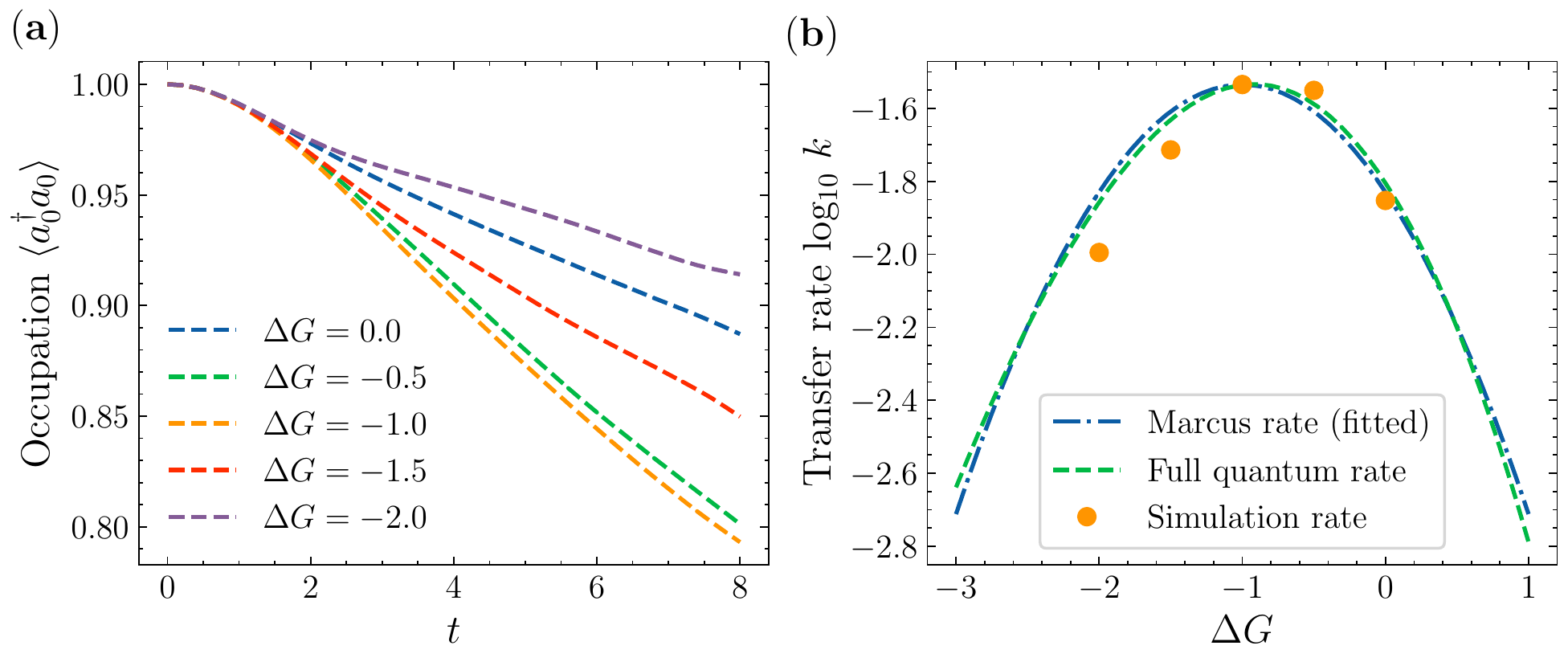}
\caption{(a) Charge occupation on the first molecule versus time at several values of $\Delta G$. (b) Charge transfer rate $k$ as a function of $\Delta G$, for $2 < t < 8$. The dots are from a linear fit based on $\braket{a^\dagger_0 a_0}(t) = -kt+c$ to the linear region in (a) and the dashed \fix{lines are the predictions of the Marcus theory (Eq.~\ref{eq:marcus}) and full quantum charge transfer theory}.}
    \label{fig:marcus}
\end{figure}

Results are shown in Fig.~\ref{fig:marcus}(a).
During an initial period ($t<2$), the reaction rate increases with $t$.
For $2 < t < 8$, the reaction reaches a steady state and the reaction rate is approximately constant.

In Fig.~\ref{fig:marcus}(b) we plot the reaction rate $k$ from $t=2$ to $t=8$.
Since the effect of temperature is not modeled in this simulation, $\beta$ is fit to a value of 2.71 by matching the simulated $k(\Delta G)$ data with the least square method.
The simulated rate reaches the maximum at $\Delta G=-1$, in agreement with the theoretical prediction.
\fix{We also include the charge transfer rate from full quantum nuclear tunneling enabled charge transfer theory~\cite{nan2009nuclear}, which predicts a very similar parabolic curve without fitting $\beta$.}

\section{Summary and Outlook}

In this paper, we introduce \tcc{}, a Python-based open-source library for quantum computational chemistry,
hosted on the GitHub repository \url{https://github.com/tencent-quantum-lab/TenCirChem}.
\tcc{} aims to provide both black-box calculations of existing quantum algorithms
and flexible interfaces for rapid prototyping
of novel computational methods.
\tcc{} features high-performance UCC calculations, noisy circuit simulation with both quantum gate error and measurement noise,
and variational quantum dynamics simulation.
\tcc{} is designed to
expose its internal data structure for inspection and modification,
enabling in-depth customization.
Although written in Python, \tcc{} reaches state-of-the-art performance
via the powerful \tc{} backend and various chemistry-focused optimizations\fix{,}  such as UCC factor expansion.
Its efficiency is demonstrated by the exact UCCSD simulation of \ce{H16} in STO-3G basis (32 qubits), \ce{H2} in cc-pVTZ basis (56 qubits), and \ce{H2O} in 6-31G(d) basis (34 qubits), requiring only moderate computational time and hardware.

The development of \tcc{} is an ongoing process.
In the future, \fix{additional QPU engine configuration options will be provided}.
Features such as 
\fix{support for higher order excitations,}
the treatment of open-shell systems,
algorithms for excited states and periodic systems,
algorithms for fault-tolerant quantum computers,
further performance optimization,
and more code examples reproducing published algorithms
are also under active consideration.

\begin{acknowledgement}
The authors thank Zhaofeng Ye for designing the \tcc{} logo.
Weitang Li would like to thank Sainan Huai and Tianqi Cai for their helpful discussions.
This work is supported by the National Natural Science Foundation of China through grant numbers 22273005 and 21788102.
This work is also supported by Shenzhen Science and Technology Program.
\end{acknowledgement}

\section*{Supporting Information}
The Supporting Information is available free of charge online, including
\begin{itemize}
    \item A brief review of UCC \fix{\cancel{ansatze}ans\"{a}tze} implemented in \tcc{}.
    \item The conventions for orbital indices and qubit indices in \tcc{}.
    \item The output of the \lstinline{print_summary} command.
    \item \fix{\cancel{The a} A}dvanced feature\fix{s} of the UCC classes.
    \item The algorithm for efficient simulation of UCC circuits.
\end{itemize}

\providecommand{\latin}[1]{#1}
\makeatletter
\providecommand{\doi}
  {\begingroup\let\do\@makeother\dospecials
  \catcode`\{=1 \catcode`\}=2 \doi@aux}
\providecommand{\doi@aux}[1]{\endgroup\texttt{#1}}
\makeatother
\providecommand*\mcitethebibliography{\thebibliography}
\csname @ifundefined\endcsname{endmcitethebibliography}
  {\let\endmcitethebibliography\endthebibliography}{}

\end{document}

\maketitle
%We don't need TOC here
%\tableofcontents

\section{Review of UCC ansatz}
We start with the general expression
\begin{equation}
\label{eq:ucc2}
    \ket{\Psi(\theta)}_{\text{UCC}} := \prod_{k=\nex{}}^1 e^{\theta_k G_k} \ket{\phi}. \
\end{equation}

\subsection{UCCSD}
For the most common case of single and double excitations, the $G_k$ has the form
\begin{equation}
\label{eq:g-single-double}
    G_k = \begin{cases}
        a^\dagger_p a_q - \text{h.c.} , \\
        a^\dagger_p a^\dagger_q a_r a_s - \text{h.c.}
        \end{cases}
\end{equation}

In this case, the disentangled UCC ansatz -- which is known as the UCCSD ansatz -- has the form
\begin{equation}
\label{eq:uccsd}
    \ket{\Psi(\theta)}_{\text{UCCSD}} := \prod_{pqrs}e^{\theta_{pqrs}(a^\dagger_p a^\dagger_q a_r a_s - \text{h.c.})}\prod_{pq}e^{\theta_{pq}(a_p^\dagger a_q - \text{h.c.})}\ket{\phi} \ .
\end{equation}
%In the remainder of this paper, we will refer to the first-order Trotterized version of the UCCSD ansatz as the the ``UCCSD ansatz".
If the number of spin-orbitals is $N$, then the number of gates in the corresponding circuit scales as $\order{N^4}$~\cite{o2019generalized}.
% if have time, explain the traditional approach in more detail, such as the scaling of the number of gates.

\subsection{$k$-UpCCGSD}
This is a variant of the UCC ansatz which satisfies both high accuracy and low ($\order{N^2}$) gate count requirements.~\cite{lee2018generalized}
Here ``G'', ``p'' and ``$k$'' stand for generalized excitation, paired double excitations, and repeating the ansatz $k$ times, respectively.
By pairing double excitations, the number of double excitations is reduced from $\order{N^4}$ to $\order{N^2}$.
More specifically, the ansatz has the form
\begin{equation}
\label{eq:kupccgsd}
    \ket{\Psi(\theta)}_{k\text{-UpCCGSD}} := \prod_{l=1}^k \prod_{pq}e^{\theta_{kpq}^{(2)}(a^\dagger_{p\alpha} a^\dagger_{p\beta} a_{q\beta} a_{q\alpha} - \text{h.c.})}\prod_{rs}e^{\theta_{krs}^{(1)}(a_r^\dagger a_s - \text{h.c.})}\ket{\phi} \ .
\end{equation}
Here $a^\dagger_{p\alpha} a^\dagger_{p\beta} a_{q\beta} a_{q\alpha}$ means spin-paired double excitations from the $q$-th spatial orbital to the $p$-th spatial orbital.

\subsection{pUCCD}
This is an efficient ansatz requiring only $\order{N}$ circuit depth and half as many qubits as other UCC ans\"{a}tze~\cite{henderson2015pair, elfving2021simulating, o2022purification, zhao2022orbital}.
pUCCD allows only paired double excitations, 
which enables one qubit to represent one spatial orbital instead of one spin orbital,
and removes the need to perform the fermion-qubit mapping.
Thus, the $\order{N^2}$ excitations can be executed on a quantum computer efficiently using a compact circuit with a linear depth of Givens-SWAP gates~\cite{elfving2021simulating}.
The Hamiltonian also takes a simpler form in this case, with only $N^2$ terms:
\begin{equation}
    H = \sum_p h_p c^\dagger_p c_p + \sum_{pq} v_{pq} c^\dagger_p c_q + \sum_{p \neq q} w_{pq} c^\dagger_p c_p c^\dagger_q c_q + E_{\rm{nuc}} \ ,
\end{equation}
where $h_p = 2h_{pp}$, $v_{pq} = (pq|pq)$ and $\omega_{pq} = 2 (pp|qq) - (pq|pq)$.
Here $p$ and $q$ are indices for spatial orbitals.
A drawback of this ansatz is its compromised accuracy, which is comparable to the doubly occupied configuration interaction (DOCI) method \cite{weinhold1967reduced} in quantum chemistry.

\section{Conventions}
\label{sec:preliminary}
In this section, we introduce the conventions used in \tcc{}, particularly for static electronic structure problems. 

\bp{Excitations. }\tcc{} uses tuples to denote unitary fermionic excitations:
\begin{itemize}
    \item $(p,q)$: denotes single excitations $a_p^\dagger a_q- a_q^\dagger a_p$
    \item $(p,q,r,s)$:  denotes double excitations $a_p^\dagger a_q^\dagger a_r a_s - a_s^\dagger a_r^\dagger a_q a_p$
\end{itemize} 
Higher-order excitations are handled similarly. e.g., a $k$-th order excitation is represented by a tuple of length $2k$, with the first half corresponding to the excitation operators, and the second half corresponding to the annihilation operators, and Hermitian conjugation is implied, e.g., $(p,q,r,s,t,u)$ denotes $a_p^\dagger a_q^\dagger a_r^\dagger a_s a_t a_u - \text{h.c.}$

\bp{Spin-orbital indexing.} Spin orbitals are indexed from 0 and ordered according to the following rules:
\begin{itemize}
    \item Beta (down) spins first, followed by alpha (up) spins
    \item Low energy orbitals first, followed by high energy orbitals
\end{itemize}

\bp{Qubit indexing.} Qubits are numbered from 0, with multi-qubit registers numbered  with the zeroth qubit on the left, e.g. $\ket{0}_{q_0}\ket{1}_{q_2}\ket{0}_{q_2}$. Unless necessary, we will omit subscripts and use a more compact notation e.g.\ $\ket{010}$ to denote multi-qubit states.  In quantum circuit diagrams qubit numbers increase downwards, starting from qubit zero at the top. Throughout, we assume the Jordan-Wigner encoding where the occupancy of spin-orbital $i$  (using the above ordering) corresponds  to the state of qubit $N - 1 - i$ where $N$ is the total number of qubits.
In other words, the qubit ordering is reversed spin-orbitals ordering.

\begin{figure}[htp]
    \centering
    \includegraphics[width=0.65\textwidth]{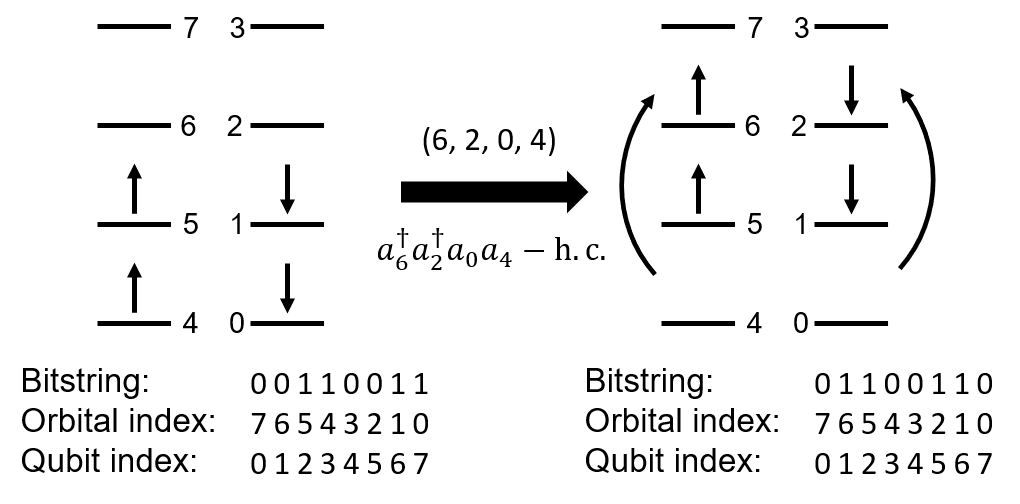}
\caption{An example of the conventions used for excitations, orbital indexing, and qubit indexing. The HF state of a 4 electron, 4 orbital system is represented as bitstring 00110011. The state is excited to 01100110 via the excitation operator (6, 2, 0, 4).}
    \label{fig:excitations}
\end{figure}

The conventions described above are summarized in Fig.~\ref{fig:excitations}, 
taking a 4-electron and 4-orbital system as an example.
Following the qubit indexing convention above, the restricted Hartree--Fock (HF) state for such a system is represented as 00110011 in bitstring form.
The highest orbital with $\alpha$-spin comes first in the bitstring, and the lowest orbital with $\beta$-spin comes last.
In quantum chemistry language, 00110011 refers to a configuration with orbitals 5, 4, 1, and 0 occupied,
while in quantum computing language, 00110011 refers to a direct product state
with the 2nd, 3rd, 6th, and 7th qubits in state $\ket{1}$ and the rest in state $\ket{0}$.
Upon application of the excitation operator (6, 2, 0, 4), the HF state transforms to 01100110.

\section{\lstinline{print_summary} output}
In this section, we provide an overview of the output by the \lstinline{print_summary} command (see Code Snippet~\ref{lst:print_summary}).  This summary is divided into a number of blocks:

\begin{lstlisting}[caption={UCCSD ansatz applied to the \ce{H_2} molecule: output of the \lstinline{print_summary} command.}, label={lst:print_summary}]
################################ Ansatz ###############################
 #qubits  #params  #excitations initial condition
       4        2             3               RHF
############################### Circuit ###############################
 #qubits  #gates  #CNOT  #multicontrol  depth #FLOP
       4      15     10              1      9  2160
############################### Energy ################################
       energy (Hartree)  error (mH) correlation energy (%)
HF            -1.116706   20.568268                 -0.000
MP2           -1.129868    7.406850                 63.989
CCSD          -1.137275   -0.000165                100.001
UCCSD         -1.137274    0.000000                100.000
FCI           -1.137274    0.000000                100.000
############################# Excitations #############################
     excitation configuration     parameter  initial guess
0        (3, 2)          1001  1.082849e-16       0.000000
1        (1, 0)          0110  1.082849e-16       0.000000
2  (1, 3, 2, 0)          1010 -1.129866e-01      -0.072608
######################### Optimization Result #########################
            e: -1.1372744055294384
          fun: array(-1.13727441)
     hess_inv: <2x2 LbfgsInvHessProduct with dtype=float64>
   init_guess: [0.0, -0.07260814651571333]
          jac: array([-9.60813938e-19, -1.11022302e-16])
      message: 'CONVERGENCE: NORM_OF_PROJECTED_GRADIENT_<=_PGTOL'
         nfev: 6
          nit: 4
         njev: 6
     opt_time: 0.02926015853881836
 staging_time: 4.5299530029296875e-06
       status: 0
      success: True
            x: array([ 1.08284918e-16, -1.12986561e-01])
\end{lstlisting}

\bp{Ansatz (lines 1-3).}  From here we see that the UCCSD ansatz is mapped (via the Jordan-Wigner transformation) to a variational quantum circuit on $4$ qubits, with two tunable parameters.  The circuit corresponds to $3$ excitation operators, which are detailed in the \lstinline{Excitations} block of the summary.  The fact that there are only two tunable parameters for three excitation operators is due to symmetry.

\bp{Circuit (lines 4-6).}  This block details the number of quantum gates that the Jordan-Wigner transformed ansatz is compiled to.  By default, \tcc{} uses the efficient circuit decomposition of \cite{yordanov2020efficient} for circuit compilation. 
The circuit depth is estimated using \qiskit{} and, as the running time of computing this estimate can be non-negligible for large circuits, the \lstinline{include_circuit=True} option
is by default False. 
The floating point operation count (FLOP) required to calculate the circuit statevector via tensor network contraction 
is estimated by the \opteinsum{} package \cite{daniel2018opt} using the default greedy contraction path-finding algorithm.

Note that, in the absence of noise, computing the output of the circuit does not require the circuit to first be decomposed explicitly into gates.  Instead, \tcc{} makes use of \textit{UCC factor expansion} (See Sec.~\ref{sec:ucc-theory}) to perform this computation more efficiently. However, if one wishes to simulate noisy quantum circuits -- corresponding, for instance, to realistic quantum hardware -- then an explicit gate decomposition is necessary. 

\bp{Energy (lines 7-13).}  This block includes the final UCCSD energy corresponding to the optimized parameter values, as well as other benchmark energies computed by \pyscf{} at Hartree--Fock, second-order Møller–Plesset perturbation theory (MP2), CCSD, and FCI levels of theory.

\bp{Excitations (lines 14-18).}  This block gives further details on the ansatz, which takes the form
\begin{align*}
e^{\theta_2 G_2}  e^{\theta_1 G_1}  e^{\theta_0 G_0}\ket{\text{HF}},
\end{align*}
where $\theta_0$ is set to be equal to $\theta_1$ due to symmetry, and where
\begin{align*}
    G_0 &= a_2^\dagger a_3 - a_3^\dagger a_2, \\
    G_1 &= a_0^\dagger a_1 - a_1^\dagger a_0, \\
    G_2 &= a_0^\dagger a_2^\dagger a_3a_1 - a_1^\dagger a_3^\dagger a_2a_0.
\end{align*}
The convention for spin-orbital indices 
is described in Sec.~\ref{sec:preliminary}.
The minimum energy found corresponds to parameter values of $\theta_0 = \theta_1 = 1.05\times 10^{-16}$ and $\theta_2 = -1.13\times 10^{-1}$.  These final values were obtained from initial guesses of $0,0, -0.0726$ respectively, which correspond to the MP2 excitation amplitudes.  The configuration bitstrings obtained after applying each excitation to $\ket{\text{HF}}$ are also given here. 
Note that the Hermitian conjugation part of the excitation 
operator annihilates $\ket{\text{HF}}$ and thus has no effect.
For comparison, the configuration bitstring corresponding to the Hartree--Fock state is 0101.

\bp{Optimization Result (lines 19-33).}  This block gives details of the procedure used to optimize the ansatz parameters, including the number of iterations required (\lstinline{nit: 3}) as well as the final energy obtained for the UCCSD ansatz (\lstinline{e: -1.137...}).

\section{The UCC classes: advanced features}
\subsection{User-specified UCC ans\"{a}tze}
UCCSD, $k$-UpCCGSD, and pUCCD are all special cases of UCC, where the excitations are restricted to take a certain form.  No setup is required for these three kinds of ansatz from users: once imported they can be used directly, e.g., via \lstinline{PUCCD(h2).kernel()}.

More general UCC ans\"{a}tze can be defined by directly specifying the corresponding excitations. 
This could be helpful, for instance, in investigating new quantum computational chemistry algorithms.  
% based on UCC, as was the case for the
% ADAPT-VQE algorithm which relies heavily on customized UCC ~\cite{grimsley2019adaptive}.
In Sec.~\ref{sec:adapt-vqe}, we will illustrate the power of this approach by implementing the ADAPT-VQE~\cite{grimsley2019adaptive} using \tcc{}, and we first provide a simpler example and implement the PUCCD ansatz from scratch (see Code Snippet~\ref{lst:custom_ucc}).
\begin{lstlisting}[language=Python, caption={Implementing the PUCCD ansatz from scratch by using a custom UCC ansatz.}, label={lst:custom_ucc}]
import numpy as np
from tencirchem import UCC, PUCCD
from tencirchem.molecule import h4

puccd = PUCCD(h4)
ucc = UCC(h4)

# only paired excitations are included
ucc.ex_ops = [
    (6, 2, 0, 4),
    (7, 3, 0, 4),
    (6, 2, 1, 5),
    (7, 3, 1, 5),
]

e1 = puccd.kernel()
# evaluate UCC energy using the PUCCD circuit parameter
e2 = ucc.energy(puccd.params)
# e1 is the same as e2
np.testing.assert_allclose(e1, e2, atol=1e-6)
\end{lstlisting}
Here, a \lstinline{UCC} class with only paired excitations is used to reproduce the \lstinline{PUCCD} class
for the \ce{H4} molecule.
The configuration is done by directly setting the \lstinline{ex_ops} attribute with tuples for the excitations.
\lstinline{param_ids} and \lstinline{initial_guess} can be set similarly 
and the full \fancylink{https://github.com/tencent-quantum-lab/TenCirChem/blob/master/example/custom_excitation.py}{Python script} is available online.
Although internally the \lstinline{UCC} class uses a quantum circuit of 8 qubits 
for simulation and the \lstinline{PUCCD} class uses only 4 qubits because of the restriction of paired excitations,
%\jon{(Comment) how can the user further set only 4 qubits to be used when using custom UCC?}
the corresponding energies with the same set of circuit parameters are exactly the same.
If custom UCC with paired excitation is desired, the \lstinline{UCC} class can be initialized with the \lstinline{hcb=True} argument.
For the index of the spin-orbitals and excitation operators, Fig.~\ref{fig:excitations} is a helpful reference.

\subsection{Active Space Approximation}
\label{sec:active-space}
As NISQ quantum hardware is limited in the number of qubits available, the active space approximation 
is frequently adopted to reduce the problem size~\cite{reiher2017elucidating, takeshita2020increasing, mizukami2020orbital, mccaskey2019quantum, o2022purification, huang2022variational}.
In many cases, inner shell molecular orbitals can be treated at the mean-field level without significant loss of accuracy, and this approximation is thus also sometimes dubbed the frozen core approximation.
Denote the set of frozen occupied spin-orbitals by $\Omega$.
The frozen core provides an effective repulsion potential $V^{\rm{eff}}$  to the remaining electrons
\begin{equation}
\label{eq:effective-pot}
    V^{\rm{eff}}_{pq} = \sum_{m \in \Omega} \left ( [mm|pq] - [mp|qm] \right).    \ 
\end{equation}
The frozen core also bears the mean-field core energy which effectively modifies
the core energy $E_{\rm{nuc}}$
\begin{equation}
    E_{\rm{core}} = E_{\rm{nuc}} + \sum_{m \in \Omega} h_{mm} + \frac{1}{2}\sum_{m, n \in \Omega} \left ( [mm|nn] - [mn|nm] \right ). \
\end{equation}
Thus, the \textit{ab initio} Hamiltonian is rewritten as
\begin{equation}
\label{eq:ham-abinit-as}
    H = \sum_{pq}(h_{pq} + V^{\rm{eff}}_{pq})a^\dagger_p a_q + \sum_{pqrs}h_{pqrs}a^\dagger_p a^\dagger_q a_r a_s + E_{\rm{core}},  \
\end{equation}
where $p$, $q$, $r$ and $s$ refer to spin-orbitals not included in $\Omega$.

%The following code performs a (2e, 2o) active space calculation of the \ce{H8} chain.
Making use of an active space approximation in \tcc{} is straightforward 
%\jon{(Comment) do we need to explain the (2e, 2o) notation?} \lc{I don't think we need to explain, it's just the common practice in chemistry}
: 
\begin{lstlisting}[language=Python, caption={UCCSD calculation of \ce{H8} with (2e, 2o) active space approximation. Note that the electron integrals in the active space are readily 
accessible via \lstinline{uccsd.int1e} and \lstinline{uccsd.int2e}.}]
from tencirchem import UCCSD
from tencirchem.molecule import h8
# (2e, 2o) active space
uccsd = UCCSD(h8, active_space=(2, 2))
uccsd.kernel()
uccsd.print_summary()
\end{lstlisting}

\subsection{Engines, Backends, and GPU}
Different engines can be specified by the \lstinline{engine} argument, e.g.,
\begin{lstlisting}[language=Python]
from tencirchem import UCCSD
from tencirchem.molecule import h4
uccsd = UCCSD(h4, engine="civector-large")
print(uccsd.kernel())
print(uccsd.energy(engine="tensornetwork"))
\end{lstlisting}
and, similar to \tc{}, computational backends can be switched at runtime as follows:
\begin{lstlisting}[language=Python]
from tencirchem import set_dtype, set_backend
set_dtype("complex64")
set_backend("cupy")
\end{lstlisting}
There are two ways to use GPUs with \tcc{}. The first is to set the backend to \lstinline{"cupy"}.
The second is to set the backend to \lstinline{"jax"} and make sure that CUDA support for \jax{} is \href{https://github.com/google/jax#installation}{properly configured}.

\subsection{Code Example: Implementing ADAPT-VQE}
\label{sec:adapt-vqe}

In this section, 
we illustrate how to use \tcc{} to build novel algorithms by implementing ADAPT-VQE~\cite{grimsley2019adaptive}.
The complete \fancylink{https://github.com/tencent-quantum-lab/TenCirChem/blob/master/docs/source/tutorial_jupyter/adapt_vqe.ipynb}{Jupyter Notebook} tutorial is available online.
The first step of the algorithm is to construct an excitation operator pool, as shown below.
Here we use all of the single and double excitations as described in Eq.~\ref{eq:g-single-double}.
Operators with the same parameter are grouped together in the operator pool so as to prevent spin contamination.
\begin{lstlisting}[language=Python, caption=Construction of an operator pool for the ADAPT-VQE algorithm.]
from tencirchem import UCC
from tencirchem.molecule import h4

ucc = UCC(h4)

# get all single and double excitations
# param_id maps operators to parameters (some operators share the same parameter)
ex1_ops, ex1_param_ids, _ = ucc.get_ex1_ops()
ex2_ops, ex2_param_ids, _ = ucc.get_ex2_ops()

# group the operators to form an operator pool
from collections import defaultdict
op_pool = defaultdict(list)
for ex1_op, ex1_id in zip(ex1_ops, ex1_param_ids):
    op_pool[(1, ex1_id)].append(ex1_op)
for ex2_op, ex2_id in zip(ex2_ops, ex2_param_ids):
    op_pool[(2, ex2_id)].append(ex2_op)
op_pool = list(op_pool.values())
\end{lstlisting}

Once the operator pool is formed, ADAPT-VQE constructs the ansatz by selecting operators in the pool iteratively, with new operators chosen to maximize the absolute energy gradient.  That is, suppose the ansatz wavefunction at a certain point in the process is $\ket{\psi}$ and the operator selected is $G_k$, then
the new ansatz, defined to be
\begin{equation}
    \ket{\Psi} = e^{\theta_k G_k} \ket{\psi}, \
\end{equation}
has a corresponding energy gradient 
\begin{equation}
    \pdv{\braket{E}}{\theta_k} = 2 \braket{\Psi|HG_k|\Psi}. \
\end{equation}
ADAPT-VQE selects $G_k$ from the operator pool such that
\begin{equation}
    \pdv{\braket{E}}{\theta_k} \bigg{|}_{\theta_k=0} = 2 \braket{\psi|HG_k|\psi} 
\end{equation}
is maximized. If multiple operators sharing the same parameter are added at the same time,
then their gradients are added together. The iterative process by which the ansatz is constructed terminates when the norm of the gradient vector falls below a predefined threshold $\epsilon$.

In the following, $\ket{\psi}$ is obtained by \lstinline{ucc.civector()} as a vector in the configuration interaction space.
\lstinline{ucc.hamiltonian(psi)} applies $H$ onto $\ket{\psi}$ and
\lstinline{ucc.apply_excitation} applies $G_k$ onto $\ket{\psi}$.

\begin{lstlisting}[language=Python, caption={Implementation of the ADAPT-VQE iteration. The complete tutorial is available online as a  \fancylink{https://github.com/tencent-quantum-lab/TenCirChem/blob/master/docs/source/tutorial_jupyter/adapt_vqe.ipynb}{Jupyter Notebook}.}]
import numpy as np

ucc.ex_ops = []
ucc.params = []
ucc.param_ids = []

MAX_ITER = 100
EPSILON = 1e-3
for i in range(MAX_ITER):
    # calculate gradient of each operator from the pool
    op_gradient_list = []
    psi = ucc.civector()
    bra = ucc.hamiltonian(psi)
    for op_list in op_pool:
        grad = bra.conj() @ ucc.apply_excitation(psi, op_list[0])
        if len(op_list) == 2:
            grad += bra.conj() @ ucc.apply_excitation(psi, op_list[1])
        op_gradient_list.append(2 * grad)
    if np.linalg.norm(op_gradient_list) < EPSILON:
        break
    chosen_op_list = op_pool[np.argmax(np.abs(op_gradient_list))]
    # update ansatz and run calculation
    ucc.ex_ops.extend(chosen_op_list)
    ucc.params = list(ucc.params) + [0]
    ucc.param_ids.extend([len(ucc.params) - 1] * len(chosen_op_list))
    ucc.init_guess = ucc.params
    ucc.kernel()
\end{lstlisting}

\section{Efficient UCC circuit simulation algorithms}
We now introduce the algorithm behind efficient UCC circuit simulation that makes \tcc{} capable of accurately simulating deep UCC circuits with more than 32 qubits.
We will first introduce the UCC factor expansion technique,  which forms the backbone of the \lstinline{"civector"} engine. 
We then move on to an efficient algorithm to evaluate energy gradients with respect to circuit parameters and other implementation details.
\subsection{UCC factor expansion}
\label{sec:ucc-theory}
In this section, we introduce the UCC factor expansion technique used to achieve efficient simulations of UCC circuits in \tcc{}.
Benchmark results are included in the main text. 
We emphasize that the
techniques described in this section are specific for efficiently simulating an ideal circuit \textit{classically}, 
but are not applicable (without large overhead costs) for real hardware or noisy circuit simulations.

\subsubsection{Traditional approach.}
Recall that the general UCC ansatz can be written as $\ket{\Psi} = \prod_{k=\nex{}}^1 e^{\theta_k G_k} \ket{\phi}$.
To compile each UCC factor $e^{\theta_k G_k}$ into a quantum circuit, 
the traditional approach is to transform
$G_k$ (in the remainder of this section we will refer to $G_k$ simply as $G$) into commutable Pauli strings,
and then simulate each Pauli string via Hamiltonian evolution.
Another approach is the YAB method described in the main text, 
which relies on multi-qubit controlled rotations.
The \lstinline{"tensornetwork"} engine follows the YAB method to simulate the UCC circuit.

\subsubsection{Factor expansion approach.}
In the more efficient \lstinline{"civector"} engine, 
we use a different approach to classically simulate the UCC circuit, which involves expanding each UCC factor into a polynomial form, implementable only on a classical computer.
While this formalism has been published multiple times in a variety of contexts~\cite{chen2021quantum, kottmann2021feasible, rubin2021fermionic}, to the best of our knowledge, \tcc{} is the first package to use such a technique for large-scale UCC circuit simulation.

In the most general case, $G$ can be written as $G = g - g^\dagger$, with
\begin{equation}\label{eq:general-G}
    g^{i_1 \dots i_m}_{j_1 \dots j_m} = a^\dagger_{i_1} \cdots a^\dagger_{i_m} a_{j_1} \cdots a_{j_m}, \
\end{equation}
where $m$ is the order of the excitation.
$g^\dagger g$ is a projector onto the space that is not annihilated by $g$~\cite{rubin2021fermionic}:
\begin{equation}
    \left (g^{i_1 \dots i_m}_{j_1 \dots j_m} \right )^\dagger g^{i_1 \dots i_m}_{j_1 \dots j_m} = (1-n_{i_1}) \cdots (1 - n_{i_m}) n_{j_1} \cdots n_{j_m}, \
\end{equation}
where $n$ is the occupation number operator.
It follows that $g^\dagger g g^\dagger = g^\dagger$.
Using this equation, together with $gg = g^\dagger g^\dagger = 0$,
it is straightforward to show that $G$ has the property
\begin{equation}
    G^3 = g^\dagger - g = -G.
\end{equation}

The corresponding UCC factor can then be expanded as
\begin{equation}
\label{eq:expanding-ucc}
\begin{aligned}
    e^{\theta G} %& = \sum_{j=0} \frac{(\theta G)^j}{j!} \\
    & = 1 + \theta G + \frac{\theta^2 G^2}{2} - \frac{\theta^3 G}{3!} 
    - \frac{\theta^4 G^2}{4!} + \frac{\theta^5 G}{5!} + \cdots \\
    & = 1 + G^2 + \sum_{j=0} \frac{(-1)^j \theta^{2j+1}}{(2j+1)!} G  
    - \sum_{j=0} \frac{(-1)^j \theta^{2j}}{(2j)!} G^2  \\
    & = 1 + \sin{\theta}G + ( 1 - \cos{\theta} ) G^2 
\end{aligned}
\end{equation}
In the special case of $G^2=-I$, the famous formula $e^{\theta G}=\cos\theta + \sin\theta G$ is recovered.
Supposing $\ket{\psi}$ is any intermediate state during circuit execution,
multiplying  $\ket{\psi}$  with Eq.~\ref{eq:expanding-ucc} yields
\begin{equation}\label{eq:ucc-poly}
    e^{\theta G} \ket{\psi} = \ket{\psi} + \sin{\theta} G \ket{\psi} + ( 1 - \cos{\theta} ) G^2  \ket{\psi}.
\end{equation}
Thus, to evaluate $e^{\theta G} \ket{\psi}$ it is sufficient to evaluate  $ G \ket{\psi} $ and $ G^2 \ket{\psi} $.

The advantages of using Eq.~\ref{eq:ucc-poly} compared to the traditional approaches are twofold.
First, Eq.~\ref{eq:expanding-ucc} significantly saves on the number of matrix multiplications required to simulate $e^{\theta G} \ket{\psi}$.
Second, as both $G$ and $G^2$ conserve particle numbers in the up and down spin sectors,
{\tcc} is able to store $\ket{\psi}$ in the particle-number conserving space instead of the whole Fock space,
just as in the standard FCI calculation.
Thus, in the following, we use configuration interaction space 
to denote this particle-number conserving sector of the Fock space. 
In traditional UCC circuit simulations,  where each quantum gate is realized by matrix multiplication,
storing wavefunction in configuration interaction space
is not possible because a single $H$ gate or $X$ gate is able to destroy the particle number conserving property.

Representing the quantum state in configuration interaction space greatly reduces the memory requirements for UCC simulation.
A closed-shell molecule with $N$ spatial orbitals and $M$ electrons has Fock space of dimension $2^{2N}$.
In contrast, the number of possible configurations in each spin sector is $C^{N}_{M/2}$, and the dimension of its configuration interaction space is only $\left ( C^{N}_{M/2} \right)^2$ .
In addition, because $G$ and $\ket{\phi}$ are both real, $\ket{\psi}$ can be represented by real numbers rather than complex numbers in classical computers. 
In Table~\ref{tab:ci-memory}, we list the memory requirements to represent wavefunctions in configuration interaction spaces and Fock spaces for several values of $N$ and $M$.
While the required memory using configuration interaction space still scales exponentially, in practice the memory saving compared with using Fock space is significant, particularly so when $|N-M|$ is large.

\begin{table}[h]
\caption{\label{tab:ci-memory}
Memory requirements to represent system wavefunctions in configuration interaction space and Fock space.
In configuration interaction space each amplitude is stored by a 64-bit floating-point number
and in Fock space each amplitude is stored by a 128-bit complex number.
}
\begin{tabular}{ccrrrrc}
\hline
\multirow{2}{*}{$N$}  & \multirow{2}{*}{$M$} & \multicolumn{2}{c}{Configuration Interaction Space} & \multicolumn{2}{c}{Fock Space}     & \multirow{2}{*}{Memory Saving}    \\
                    &                    & \multicolumn{1}{c}{Dimension} & \multicolumn{1}{c}{Memory}  & \multicolumn{1}{c}{Dimension} & \multicolumn{1}{c}{Memory} & \\ 
                    \hline
2 & 2 & 4 & 32 B & 16 & 256 B & 8x \\
\multirow{2}{*}{4}  & 2    &   16    &   128 B & \multirow{2}{*}{256}  &  \multirow{2}{*}{4.1 kB} &  32x   \\
                    & 4    &   36     &   288 B   &        &    &   14x  \\ 
\multirow{2}{*}{8}  & 4    &   784    &   6.3 kB & \multirow{2}{*}{65,536}  &  \multirow{2}{*}{1.0 MB} &  167x   \\
                    & 8    &   4,900     &   39.2 kB   &        &    &   27x  \\ 
\multirow{2}{*}{16}  & 8    &   3,312,400 &   26.5 MB & \multirow{2}{*}{4,294,967,296}  &  \multirow{2}{*}{68.7 GB} &  2,593x   \\
                    & 16    &   165,636,900  &   1.3 GB   &        &    &   52x  \\ 
48 & 4 & 1,272,384 & 10.1 MB &$ 7.9 \times 10^{28}$ & 1.3 QB &  $1.2\times 10^{23}$x \\
\hline
\end{tabular}
\end{table}

\subsection{Gradients with respect to circuit parameters}
\label{sec:ucc-gradient}
\tcc{} uses an efficient algorithm to calculate the energy gradient with respect to the parameters~\cite{luo2020yao}.
The algorithm is applicable to the \lstinline{"civector"} engine, and for the \lstinline{"tensornetwork"} engine traditional auto-differentiation is implemented.
For $j=1,2,\ldots, N_{ex}$ define
\begin{equation}
    \ket{\psi_j} = \prod_{k=j}^1 e^{\theta_k G_k} \ket{\phi}, \
\end{equation}
and  $\ket{\psi'_j}$
\begin{equation}
    \langle \psi'_j |  = \langle \psi | \hat H \prod_{k=\nex}^{j} e^{\theta_k G_k}, \
\end{equation}
where $H$ is the system Hamiltonian.
The energy expectation value can be written as
\begin{equation}
    \langle E \rangle = \langle \psi'_{j+1} | \psi_j \rangle, \
\end{equation}
and the energy gradient as
\begin{equation}
\label{eq:grad-numerical}
        \frac{\partial \langle E \rangle}{\partial \theta_j} =
    2\langle \psi'_{j+1} |  G_j | \psi_j \rangle. \
\end{equation}

Once $| \psi \rangle$ is obtained, $\langle \psi^{(1)}_{\nex+1} |=\langle \psi | \hat H$
and $| \psi_{\nex} \rangle = |\psi \rangle$
are calculated. The remaining $\langle \psi'_j |$ and $| \psi_{j-1} \rangle$ are then obtained by the recurrence relation
\begin{equation}
    \begin{aligned}
    \langle \psi'_j | & = \langle \psi'_{j+1} | e^{\theta_j G_j}, \ \\
    | \psi_{j-1} \rangle & = e^{-\theta_{j} G_{j}}  | \psi_{j} \rangle, \
    \end{aligned}
\end{equation}
and, for each $\langle \psi'_{j} |$ and $| \psi_{j-1} \rangle$ pair,
$\frac{\partial \langle E \rangle}{\partial \theta_j}$ is evaluated by Eq.~\ref{eq:grad-numerical}.
This algorithm thus computes all gradients in one iteration over $j$, 
requiring only a constant amount of memory.

\subsection{Other techniques for efficient simulation}

With UCC factor expansion and efficient gradient computation, \tcc{} is able to simulate much larger molecular systems compared to traditional simulation packages. 
% In general, the computational cost is essentially at the same level as FCI calculations \jon{(Comment) any justification for this claim? Isn't FCI very expensive? If we don't beat FCI why not just use FCI?}.
\tcc{} also use several other techniques to accelerate calculations. In particular (i)
  initial values of $\theta_k$ are set to the corresponding $t_2$ amplitudes obtained by MP2, to enable faster convergence; (ii)
%For closed-shell molecules, configurations with spin-flip symmetry look, when ask also ask ci space
double excitation operators with $t_2$ very close to zero are screened out by default
because the excitation is likely prohibited by molecular point-group symmetry~\cite{cao2022progress}; (iii)
double excitation operators are also sorted by $t_2$ amplitudes 
to avoid ambiguous ans\"{a}tze~\cite{grimsley2019trotterized}.

\bibliography{refs}